\documentclass[aps,prd,amsmath,amssymb,preprintnumbers,nofootinbib]{revtex4-1}
\usepackage[utf8]{inputenc} 
\usepackage{graphicx}

\usepackage{dcolumn}
\graphicspath{{./figures/}}
\usepackage{url}
\usepackage[bookmarks, pagebackref=false]{hyperref}
\usepackage[usenames,dvipsnames]{xcolor}
\definecolor{orange}{cmyk}{0,0.5,1,0}
\definecolor{rossoCP3}{cmyk}{0,.88,.77,.40}
\definecolor{graa}{rgb}{0.8,0.8,0.8}
\definecolor{blaa}{rgb}{0.2,0.2,0.6}
		\hypersetup{
			colorlinks,
			bookmarksopen,
			bookmarksnumbered,
			citecolor=blaa, 		
			linkcolor=rossoCP3,	
			urlcolor=rossoCP3,			
			}
\usepackage{amsthm}
\usepackage{bm}
\usepackage{bbm}
\usepackage{pxfonts}

\usepackage{amsmath,amssymb,amsfonts}
\usepackage{color}
\usepackage{float}
\usepackage{hyperref}
\usepackage[Symbolsmallscale]{upgreek}
\usepackage{amsmath}
\usepackage{amsfonts}
\usepackage{amssymb,dsfont}
\usepackage{graphicx}
\usepackage{amssymb}
\usepackage[vcentermath]{youngtab}
\usepackage[all]{xy}
\usepackage{pstricks}
\usepackage{dsfont}%
\usepackage{bbold}
\usepackage{subcaption}

\usepackage{placeins}
\usepackage{xspace}
\usepackage{cancel} 

\usepackage{slashed}

\usepackage{natbib}

\usepackage{feynmf}

\usepackage{braket}

\newcommand{\beq}{\begin{eqnarray}}
\newcommand{\eeq}{\end{eqnarray}}

\newcommand{\bmp}{\noindent\begin{minipage}{16cm}}
\newcommand{\emp}{\end{minipage}\vskip 7mm} 

\newcommand   \cO {\mathcal{O}}

\newcommand{\Tr}{\text{Tr}}
\newcommand*{\del}{\mathop{\mathrm{{}\partial}}\mathopen{}}

\def\lsim{\mathrel{\rlap{\lower4pt\hbox{\hskip1pt$\sim$}}
    \raise1pt\hbox{$<$}}}                
\def\gsim{\mathrel{\rlap{\lower4pt\hbox{\hskip1pt$\sim$}}
    \raise1pt\hbox{$>$}}}                

\newcommand{\be}{\begin{equation}} \newcommand{\ee}{\end{equation}}
\newcommand{\bea}{\begin{eqnarray}}  \newcommand{\eea}{\end{eqnarray}}

\newcommand   \cA {\mathcal{A}}

\begin{document}

\title{More on the Weak Gravity Conjecture via Convexity of Charged Operators
  }
\author{Oleg {\sc Antipin}$^{\color{rossoCP3}{\clubsuit}}$}
\email{oantipin@irb.hr}
\author{Jahmall {\sc Bersini}
$^{\color{rossoCP3}{\clubsuit}}$}
\email{jbersini@irb.hr}
\author{Francesco {\sc Sannino} $^{\color{rossoCP3}{\varheartsuit ,\diamondsuit},\color{rossoCP3}{\heartsuit}}$}
\email{sannino@cp3.sdu.dk}
\author{Zhi-Wei Wang $^{\color{rossoCP3}{\diamondsuit}}$}
\email[Corresponding author:\,]{wang@cp3.sdu.dk}
\author{Chen Zhang $^{\color{rossoCP3}{\spadesuit}}$}
\email[Corresponding author:\,]{chen.zhang@fi.infn.it}
\affiliation{{ $^{\color{rossoCP3}{\clubsuit}}$ Rudjer Boskovic Institute, Division of Theoretical Physics, Bijeni\v cka 54, 10000 Zagreb, Croatia}\\
{$^{\color{rossoCP3}{\varheartsuit}}$Scuola Superiore Meridionale, Largo S. Marcellino, 10, 80138 Napoli NA, Italy.}\\{
$^{\color{rossoCP3}{\diamondsuit}}$\color{rossoCP3} {CP}$^{ \bf 3}${-Origins}} \& the Danish Institute for Advanced Study {\color{rossoCP3}\rm{Danish IAS}},  University of Southern Denmark, Campusvej 55, DK-5230 Odense M, Denmark. \\\mbox{ $^{\color{rossoCP3}{\heartsuit}}$Dipartimento di Fisica “E. Pancini”, Università di Napoli Federico II | INFN sezione di Napoli}\\ \mbox{Complesso Universitario di Monte S. Angelo Edificio 6, via Cintia, 80126 Napoli, Italy.}\\
{$^{\color{rossoCP3}{\spadesuit}}$ INFN Sezione di Firenze, Via G. Sansone 1, I-50019 Sesto Fiorentino, Italy}}

\begin{abstract}
The Weak Gravity Conjecture  has {recently} been re-formulated in terms of a particle with non-negative self-binding energy.  Because of the dual conformal field theory (CFT) formulation in the anti-de Sitter space, the conformal dimension $\Delta (Q)$ of the lowest-dimension operator with charge $Q$ under some global $U(1)$ symmetry  must be a convex function of $Q$. This property has been conjectured to hold  for any (unitary) conformal field theory and generalized to larger global symmetry groups. Here we refine and further test the convex charge conjecture via semiclassical computations for fixed charge sectors of different theories in various dimensions.
We analyze the convexity properties of the leading and next-to-leading order terms stemming from the semiclassical computation, de facto, extending previous tests beyond the leading perturbative contributions and to arbitrary charges. In particular, the leading contribution is sufficient to test convexity in the semiclassical computations.
 {We also consider intriguing cases in which the models feature a transition from real to complex conformal dimensions either as a function of the charge or number of matter fields. As a relevant example of the first kind, we investigate the $O(N)$ model in $4+\epsilon$ dimensions. } As an example of the second type, we consider
the $U(N)\times U(M)$ model in $4-\epsilon$ dimensions. Both models display a rich dynamics where, by changing the number of matter fields and/or charge, one can achieve dramatically different physical regimes. We discover that whenever a complex conformal dimension appears, the real part   satisfies the convexity property. \\
 {\footnotesize  \it Preprint: RBI-ThPhys-2021-35}

\end{abstract}

\maketitle

\tableofcontents

\section{Introduction}

Although there is no consensus on the ultimate theory of quantum gravity there are efforts in understanding the overall  properties  that a consistent theory of quantum gravity and particle physics  should respect.  Inspired by string theory the Swampland program proposed in \cite{Vafa:2005ui} and nicely summarized in  \cite{Palti:2019pca,vanBeest:2021lhn} aims, indeed, at providing relevant constraints on low energy effective theories from their consistent embedding in a well defined quantum gravity theory.
It is fair to say that some conjectures are more general than others. Nevertheless crispier definitions that can help prove the conjectures or provide wider tests, beyond  string theory are desirable.

Inspired by the weak gravity conjecture, originally formulated in~\cite{Arkani-Hamed:2006emk}, according to which, at low energy, gravity is the weakest force, Aharony and Palti~\cite{Aharony:2021mpc}  introduced a  further conjecture  relating it to possible convexity of the conformal dimensions of certain fixed charge operators in conformal field theories (CFT)s. This convex charge conjecture can be viewed as the holographic dual of the positive binding energy conjecture in AdS space also proposed in~\cite{Aharony:2021mpc}. This claims the existence of a charged particle with a non-negative self-binding energy for a gauge theory coupled to gravity~\footnote{The positive binding conjecture is motivated from a version of the weak gravity conjecture called the repulsive force conjecture~\cite{Arkani-Hamed:2006emk,Palti:2017elp}, whose relation to the original weak gravity conjecture has been clarified in ~\cite{Heidenreich:2019zkl}. For different versions of the weak gravity conjecture and relations between them, we refer the reader to ref.~\cite{Cheung:2014vva,Nakayama:2015hga,Heidenreich:2016aqi,Palti:2017elp,
Lust:2017wrl,Andriolo:2018lvp,Lee:2018urn,Lee:2018spm,Heidenreich:2019zkl}.}. Moreover, the convex charge conjecture is suggested to hold for general CFTs and it has been extensively tested~\cite{Aharony:2021mpc} via perturbation theory, large $N$ and semiclassical expansions. Intriguingly, the CFTs might not even feature gravitational duals.

The overarching goal of our work is to refine and test the convex charge conjecture via semiclassical computations for unitary and non-unitary CFTs in different space-time dimensions within and beyond ordinary perturbation theory.
This is possible thanks to  advances in studying CFTs with continuous global symmetries in the presence of a  conserved charge $Q$.  The large charge dynamics was first investigated in \cite{Hellerman:2015nra, Alvarez-Gaume:2016vff,Monin:2016jmo,Jafferis:2017zna, Hellerman:2017sur} where one can  extract  the relevant conformal dimensions in inverse powers of the charge as reviewed \cite{Gaume:2020bmp}.   One can deform the CFT    \cite{Orlando:2019skh, Orlando:2020yii} to acquire novel information about the spectrum and dynamics of near conformal dynamics relevant for phenomenological applications \cite{Sannino:2009za, Cacciapaglia:2020kgq}.  The methodology employed to investigate all of the above
is often referred to as semiclassical in the sense that the path-integral is dominated by trajectories near the classical solution of the equation of motions. Another relevant limit is the one in which the CFT is perturbative and controlled by a small parameter $\epsilon$. The latter can emerge because there is a non-trivial interacting  Banks-Zaks type \cite{Banks:1981nn} fixed point near the loss of asymptotic freedom of either perturbatively safe \cite{Litim:2014uca} or infrared nature. The safe case was investigated first in \cite{Orlando:2019hte}. Another way to introduce a small parameter is to slightly modify the number of space-time dimensions. This typically endows, for UV free theories, perturbative infrared fixed points. Interestingly, the charge expansion captures higher orders in the ordinary perturbative coupling corrections \cite{Badel:2019oxl, Arias-Tamargo:2019xld, Antipin:2020abu, Antipin:2020rdw,Jack:2020wvs}. The reason being that the presence of a small parameter allows studying the fixed-charge sectors of a CFT by defining a 't Hooft-like coupling $\mathcal{A} = \epsilon  Q$ in which one can take the limit $\epsilon \to 0$ while maintaining $\mathcal{A}$ fixed. In fact, one can  resum the ordinary perturbation series by providing all-order results in the $\mathcal{A}$ coupling.
Below, we will introduce the models which will be used in this work.
The $O(N)$ model was first investigated via semiclassical methods for any $N$  in $4-\epsilon$ dimensions in \cite{ Antipin:2020abu}. The results were  successfully tested against ordinary perturbation theory in \cite{Jack:2021ypd} to four loops. Later on the $O(N)$ model was also investigated via the  semiclassical approach at large $N$ in various dimensions in \cite{Arias-Tamargo:2020fow,Giombi:2020enj}. The $O(N)$ model in $3-\epsilon$ dimensions was investigated~\cite{Badel:2019khk,Jack:2020wvs}. The study of the quite involved  $U(N)\times U(M)$ model in $4-\epsilon$ dimensions appeared in~\cite{Antipin:2020rdw,Antipin:2021akb}. The equivalence between $O(N)$ quartic and cubic model near six dimensions was firstly studied in \cite{Giombi:2020enj} and further investigated within the fixed charge sector in~\cite{Antipin:2021jiw}. 

For the models above we analyze the convexity properties of the leading and next-to-leading order terms stemming from the semiclassical computation.   The  interesting feature of the $O(N)$ model in $4+\epsilon$ dimensions is that a complex conformal dimension develops when increasing the charge.  On the other hand for
the $U(N)\times U(M)$ model in $4-\epsilon$ dimensions the emergence of complex conformal dimensions is function of the number of matter fields. Both models describe, depending on the parameters, dramatically different physical regimes. We discover that whenever a complex conformal dimension arises, the real part   satisfies the convexity property.

We organize our paper as follows.
In section II, we clarify a few conceptual issues about formulations of convex charge conjectures and emphasize the relevance of semiclassical methods for testing the conjecture.
In section III, we carry out various tests for the conjecture by using the semiclassical method. We start with $O(N)$ model in $4-\epsilon,\,4+\epsilon,\,3-\epsilon$ dimensions and $2<d<4$ dimension. Then we move on to the $U(N)\times U(M)$ model in $4-\epsilon$ dimensions.
We conclude the main results of our work in section IV.


\section{Formulations of the Conjecture: Refinements and the Use of Semiclassics}

\subsection{Mathematical Preparation}

The convex charge conjecture proposed in ref.~\cite{Aharony:2021mpc} involves functional inequalities of the form
\begin{align}
f(x+y)\geq f(x)+f(y)
\label{eq:sadef}
\end{align}
for $x,y\in\mathbb{R}$ and a function $f:\mathbb{R}\rightarrow\mathbb{R}$. Mathematically a function $f$ satisfying
Eq.~\ref{eq:sadef} is called \emph{superadditive}~\cite{Kuczma:2000fe}. We note that this property of superadditivity is technically different~\footnote{This difference has been pointed out in footnote 4 of ref.~\cite{Aharony:2021mpc}. Here we give more details.} from the usual definition of functional \emph{convexity} which in the one-variable case is equivalent to requiring
\begin{align}
f(\lambda x+(1-\lambda)y)\leq \lambda f(x)+(1-\lambda)f(y)
\end{align}
for a continuous function $f$ and all $x,y$ belonging to a convex subset in $\mathbb{R}$, and all $\lambda\in [0,1]$. The relation between superadditivity and convexity is embodied in the following theorem due to M. Petrovic ~\cite{Kuczma:2000fe}.

\textbf{Theorem (Petrovic)}$\quad$ Let $[0,a)\subset\mathbb{R},0<a\leq\infty$, and let $f:[0,a)\rightarrow\mathbb{R}$ be a continuous and convex function. Then for every $n\in\mathbb{N}$ and every $x_1,x_2,...,x_n\in[0,a)$ such that $x_1+x_2+...+x_n\in[0,a)$ we have
\begin{align}
f(x_1+x_2+...+x_n)\geq f(x_1)+...+f(x_n)-(n-1)f(0)\,.
\label{eq:petrovic}
\end{align}
The proof of the theorem can be found in ref.~\cite{Kuczma:2000fe}. To make contact with the convex charge conjecture we may identify $f(Q)$ as the scaling dimension as a function of charge $Q$~\footnote{Note that the charge $Q$ is originally only constrained to be integers. On the other hand, most known computational methods deliver $f(Q)$ as a function of a continuous variable $Q$, which is subject to tests of convexity.}. It is then obvious that we always have
\begin{align}
f(0)=0
\end{align}
and in the $n=2$ case Eq.~\eqref{eq:petrovic} becomes Eq.~\eqref{eq:sadef}. With the conditions of the theorem in mind, the consequence of the theorem can be simply put as ``convexity implies superadditivity''. The converse need not be true. In this work mostly we will directly test the convexity of the functions involved and use the convexity to infer the superadditivity.

One can test the convexity of a function using the following criterion: if a function $f(Q)$ is twice differentiable in an open interval $J\subset\mathbb{R}$, then $f(Q)$ is convex if and only if $f''(Q)$ is nonnegative in $J$~\cite{Kuczma:2000fe}. In using this criterion, one should note that $f(Q)$ is required to be twice differentiable in the whole open interval $J$. If this is not satisfied, for example due to some kink, then this criterion is not applicable. When we examine the real part of the scaling dimension of fixed-charge operators in certain non-unitary theories we do expect such kinks at critical values of the charge. In such cases, one should resort to definitions at a more basic level to determine whether the functions involved are convex.

\subsection{Formulations of the Convex Charge Conjecture}

As a preparation for testing the convex charge conjecture in generic situations, we will introduce several formulations of the convex charge conjecture. They are common in spirit but different in technicalities.
We are not aware of arguments leading to proof of their equivalence or inequivalence. Nevertheless, we find it instructive to state them clearly, show their distinctions, and keep in mind which version is tested in each case
of interest.

To this end let us first introduce some notations. Suppose we consider a conformal field theory (CFT) with a continuous internal global symmetry group $\mathcal{G}$~\footnote{Generalization to the cases in which there are additional discrete internal symmetry group is straightforward.}. Consider the charge space $V$ of $\mathcal{G}$,
which is the real vector space spanned by all fundamental weights of $\mathcal{G}$. Any weight $w$ of an irreducible representation of $\mathcal{G}$ is an element of $V$. For any given weight $w$, there must be a multiplet (called $M$) of operators which transforms in an irreducible representation $\Gamma_M$ of $G$ such that $\Gamma_M$ contains $w$ as its weight and which has the lowest possible scaling dimensions among all such multiplets. We call $\Gamma_M$ the lowest-lying representation of the given weight $w$, and denote it as
\begin{align}
r_L[w]\equiv\Gamma_M\,.
\end{align}
We then introduce the following notations.
\begin{enumerate}
\item The highest weight of some irreducible representation $r$ will be denoted $w_h[r]$.
\item The irreducible representation corresponding to some highest weight $w$ will be denoted $r_H[w]$.
\item The scaling dimension of the lowest dimension operator in the irreducible
representation $r$ will be denoted $\Delta(r)$.
\end{enumerate}

Note that $\Delta(r_L[w])$ is just the scaling dimension of the multiplet $M$ introduced above in the definition of the lowest-lying representation for a given weight $w$.

Suppose $r_R$ is a reducible representation of $\mathcal{G}$ which has the decomposition
\begin{align}
r_R=\underset{i}{\mathop{\oplus }}\,{{m}_{i}}{{r}_{i}}\,,
\end{align}
where $r_i$'s are irreducible representations with multiplicity $m_i$. $w$ is a given weight of $r_R$. Then we introduce the notation
\begin{align}
\bar{\Delta }(r_R;w)\equiv \min \Big\{\Delta ({{r}_{i}})|w\text{ belongs to }{{r}_{i}}\Big\}\,.
\end{align}

We will also need the notion of a \emph{rational direction}. A rational direction in the charge space $V$ associated with the Lie group $\mathcal{G}$ is a ray from the origin which intersects another lattice point (i.e. weight)~\cite{Heidenreich:2019zkl}. Obviously each nonzero weight specifies a unique rational direction and the set of rational directions is dense within the set of all directions. For any rational direction $R$, suppose the distance from the origin to the nearest nonzero weight in the rational direction is $d_0$, then points along $R$ at distance $nd_0$ from the origin ($n$ is any nonnegative integer) are also weights. Conversely, any weight along $R$ has a distance $d$ to the origin whose ratio to $d_0$ is some nonnegative integer:
$d=nd_0,n\in\mathbb{N}$. This integer $n$ of a given weight $w$ will be called its distance index.

We are now in a place to state more precisely the convex charge conjecture of Aharony and Palti~\cite{Aharony:2021mpc}.

\textbf{Convex Charge Conjecture (Aharony \& Palti)}$\quad$ Consider any unitary CFT with a continuous internal global symmetry group $\mathcal{G}$, and any rational direction $R$ of $\mathcal{G}$, then there exists a weight $w_0$ in $R$ such that if we define a real number $\tilde{\Delta}(Q)$ for any nonnegative integer $Q$ as follows
\begin{align}
\tilde{\Delta }(Q)\equiv \Delta ({{r}_{H}}[Q{{w}_{h}}[{{r}_{L}}[{{w}_{0}}]]])\,.
\label{eq:dtilde}
\end{align}
Then $\tilde{\Delta}(Q)$ satisfies the following superadditive property for all nonnegative integers $Q_1,Q_2$
\begin{align}
\tilde{\Delta }({{Q}_{1}}+{{Q}_{2}})\ge \tilde{\Delta }({{Q}_{1}})+\tilde{\Delta }({{Q}_{2}})\,.
\end{align}
Moreover, the distance index of $w_0$ is of $\mathcal{O}(1)$.

In this version of the conjecture (called original version hereafter), the choice of the irreducible representation $r_H$ is in accordance with the footnote 1 of ref.~\cite{Aharony:2021mpc}.

There is a second version of the conjecture which is motivated from an operator product expansion (OPE) viewpoint (and has been alluded to in ref.~\cite{Aharony:2021mpc}), which we now state as follows (and called OPE version hereafter).

\textbf{Convex Charge Conjecture (OPE)}$\quad$ Consider any unitary CFT with a continuous internal global symmetry group $\mathcal{G}$, and any rational direction $R$ of $\mathcal{G}$, then there exists a weight $w_0$ in $R$ such that if we define a real number $\bar{\Delta}(Q)$ for any nonnegative integer $Q$ as follows~\footnote{$\text{Sy}{{\text{m}}^{Q}}({r})$ denotes the symmetric product of $Q$ copies of the representation $r$.}
\begin{align}
\bar{\Delta }(Q)\equiv \bar{\Delta }\text{(Sy}{{\text{m}}^{Q}}({{r}_{L}}[{{w}_{0}}]);Q{{w}_{0}})\,.
\end{align}
Then $\bar{\Delta}(Q)$ satisfies the following superadditive property for all nonnegative integers $Q_1,Q_2$
\begin{align}
\bar{\Delta }({{Q}_{1}}+{{Q}_{2}})\ge \bar{\Delta }({{Q}_{1}})+\bar{\Delta }({{Q}_{2}})\,.
\end{align}
Moreover, the distance index of $w_0$ is of $\mathcal{O}(1)$.

A third version of the conjecture is motivated by certain applications of fixed-charge semiclassical methods in which one maps the CFT to a cylinder using Weyl invariance. In these applications weights play a more prominent role than irreducible representations. Thus we will call this version the weight version hereafter and state it as follows.

\textbf{Convex Charge Conjecture (Weight)}$\quad$ Consider any unitary CFT with a continuous internal global symmetry group $\mathcal{G}$, and any rational direction $R$ of $\mathcal{G}$, then there exists a weight $w_0$ in $R$ such that if we define a real number $\hat{\Delta}(Q)$ for any nonnegative integer $Q$ as follows
\begin{align}
\hat{\Delta }(Q)\equiv \Delta ({{r}_{L}}[Q{{w}_{0}}])\,.
\end{align}
Then $\hat{\Delta}(Q)$ satisfies the following superadditive property for all nonnegative integers $Q_1,Q_2$
\begin{align}
\hat{\Delta }({{Q}_{1}}+{{Q}_{2}})\ge \hat{\Delta }({{Q}_{1}})+\hat{\Delta }({{Q}_{2}})\,.
\label{key_equation}
\end{align}
Moreover, the distance index of $w_0$ is of $\mathcal{O}(1)$.

A few remarks are in order regarding the relations between these three versions of the conjecture. If $\mathcal{G}$ is Abelian (e.g. $U(1)$), then three versions of the conjecture are manifestly equivalent. Complications may arise if $\mathcal{G}$ contains non-Abelian factor(s). The OPE version is conceptually different from the original version because $\text{Sy}{{\text{m}}^{Q}}({{r}_{L}})$ can be a reducible representation
while $r_H$ in Eq.~\eqref{eq:dtilde} is just one irreducible component in the decomposition of $\text{Sy}{{\text{m}}^{Q}}({{r}_{L}})$ which contains the weight $Qw_0$. In the OPE version different irreducible components in the decomposition of $\text{Sy}{{\text{m}}^{Q}}({{r}_{L}})$ must compete to determine which one has the lowest scaling dimension and thus is responsible for the evaluation of $\bar{\Delta}(Q)$. However, if for the given CFT a weight $w_0$ in the rational direction $R$ of its symmetry group lies in the Weyl group orbit of $w_h[r_L[w_0]]$, then the OPE version and the original version of the conjecture are equivalent in the rational direction $R$. This is because in such a case the decomposition of $\text{Sy}{{\text{m}}^{Q}}({{r}_{L}})$ contains a unique irreducible representation (which is $r_H$ in Eq.~\eqref{eq:dtilde}) that contains $Qw_0$. The uniqueness originates from the fact that $Qw_0$ is equivalent to the highest weight (up to Weyl group transformations).

For the weight version of the conjecture, all irreducible representations that contain $Qw_0$ as a weight must compete to determine which one has the lowest scaling dimension and thus is responsible for the evaluation of $\hat{\Delta}(Q)$. Only when all such irreducible representations belong to the decomposition of $\text{Sy}{{\text{m}}^{Q}}({{r}_{L}}[{{w}_{0}}])$ can we conclude that the weight version and the OPE version of the conjecture are equivalent. One possibility for this to occur is when for all nonnegative integer $Q$, $Qw_0$ lies in the Weyl group orbit of $w_h[r_L[Qw_0]]$; in such a case all three versions of the conjecture are equivalent.

\begin{figure}
\begin{center}
\includegraphics[width=0.7\textwidth]{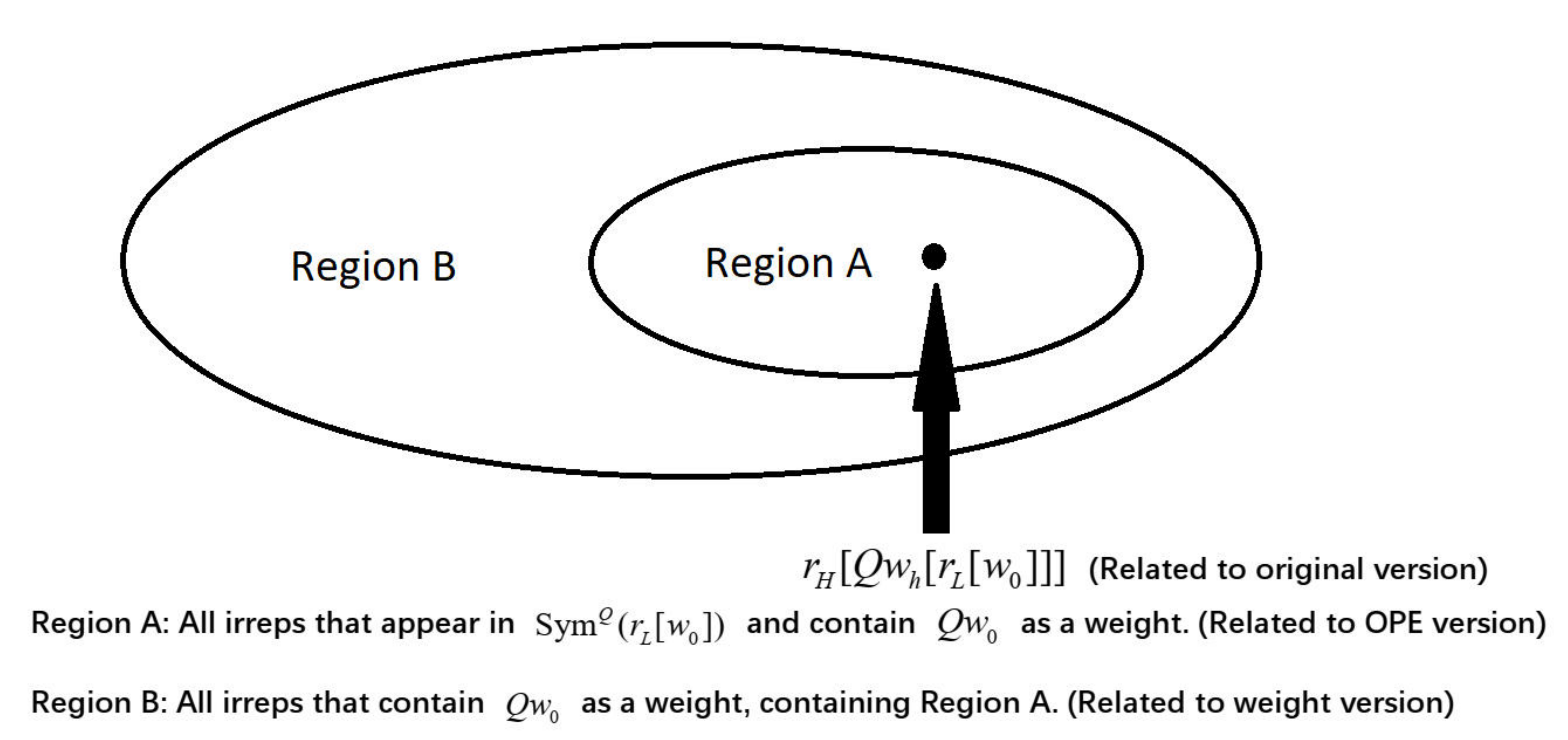}
\caption{\label{fig:wgcversions} The relations between three versions of the convex charge conjecture as illustrated via the set of irreducible representations of operator multiplets which participate in the competition for the lowest scaling dimension. For the original version of the conjecture, only one irreducible representation is involved, which is ${{r}_{H}}[Q{{w}_{h}}[{{r}_{L}}[{{w}_{0}}]]]$ and represented by the black dot. For the OPE version of the conjecture, all irreducible representations that appear in the decomposition of $\text{Sy}{{\text{m}}^{Q}}({{r}_{L}})$ and contain $Qw_0$ as a weight participate (corresponding to region A), including ${{r}_{H}}[Q{{w}_{h}}[{{r}_{L}}[{{w}_{0}}]]]$. For the weight version of the conjecture, all irreducible representations that contain $Qw_0$ as a weight participate (corresponding to region A which contains region B as its subset).}
\end{center}
\end{figure}

The relations between three versions of the conjecture are illustrated in Fig.~\ref{fig:wgcversions}. In this figure the three versions are distinguished by the set of irreducible representations that participate in the competition for the lowest scaling dimension. Under certain circumstances it might occur that region B collapses to region A, or region A collapses to the black dot, leading to the equivalence between corresponding version in those cases, as discussed above.

The underlying reason behind this complication from three versions of the conjecture is that generically it is a difficult task to determine the lowest-lying representation (with or without further constraints) for a given weight,
except for simple enough cases. Attempts have been made in ref.~\cite{Antipin:2021akb} for operators in a variety of representations of the critical $U(N)\times U(M)$ Higgs theory but it would be very complicated to generalize further.
Lacking a proof of their equivalence or inequivalence, what can be concluded currently is that the conjecture that is being tested directly via semiclassical computation through a Weyl map in Section~\ref{sec:oNmodel_4-epsilon}~\ref{subsec:3b}~\ref{sec:u1model3e}~\ref{subsec:unumtest} corresponds to the weight version. For the cases studied in Section~\ref{sec:oNmodel_4-epsilon}~\ref{subsec:3b}~\ref{sec:u1model3e}
$w_0$ can be taken to be the highest weight of the fundamental representation of $O(N)$ and one has
$w_h[r_L[w_0]]=w_0$. If we further assume the no-level-crossing condition $r_H[Qw_0]=r_L[Qw_0]$ to hold in the $O(N)$ models for every nonnegative integer $Q$, then the three versions of the conjecture are also equivalent. On the other hand Section~\ref{sec:onmodelgend} is based on semiclassical large $N$ results obtained by operator insertion directly, without a Weyl map. The version of the conjecture that is being tested for $O(N)$ model corresponds to the OPE version which is also equivalent to the original version in this case. Section ~\ref{subsec:unumtest} studies several charge configurations in the $U(N)\times U(M)$ models, leading to a situation that is more obscured as to the relation between three versions of the conjecture, which we leave for future work to explore.

Finally, all three versions of the conjectures above are about one rational direction. Thus in the resulting superadditivity inequality the charge can be represented as a nonnegative integer. One natural generalization is to consider a superadditivity inequality in which the charges are represented as vectors restricted to some regions. In the case of a one-dimensional charge space the charges are restricted to be nonnegative. In multi-dimensional cases the natural generalization would be to consider fixed-sign charge space, that is the space spanned by Cartan generators which have fixed signs along diagonal entries. For example, in the case of $SU(3)$, Cartan generators can be represented as $3\times 3$ traceless diagonal matrices and thus there are six fixed-sign charge spaces, corresponding to
\begin{align}
(+,+,-),(+,-,+),(-,+,+),(-,-,+),(-,+,-),(+,-,-)\,.
\end{align}
A zero entry belongs to both signs. The traceless condition makes $(+,+,+),(-,-,-)$ charge spaces trivial.

It is thus natural to formulate a generalized convex charge conjecture as follows.

\textbf{Convex Charge Conjecture (Generalized, falsified)}$\quad$ Consider any unitary CFT with a continuous internal global symmetry group $\mathcal{G}$, then for any two nonzero weights $w_1$ and $w_2$ in a fixed-sign charge space of $\mathcal{G}$, there exists an $\mathcal{O}(1)$ positive integer $n_0$ such that for any nonnegative integer $Q$ we have
\begin{align} \label{genconvex}
\Delta ({{r}_{L}}[Q{{n}_{0}}({{w}_{1}}+{{w}_{2}})])\ge \Delta ({{r}_{L}}[Q{{n}_{0}}{{w}_{1}}])+\Delta ({{r}_{L}}[Q{{n}_{0}}{{w}_{2}}])\,.
\end{align}

However we found examples that violate this generalized version of convex charge conjecture, which we discuss in Section~\ref{subsec:unumtest}.

\subsection{The Use of Semiclassics}

At small values of the charge, conventional perturbation theory is valid. The conventional perturbation for the scaling dimensions $\Delta(Q)$ of fixed-charge operators is organized as a multi-variable Taylor expansion in couplings, with coefficients being polynomials of the charge. The degrees of the polynomials are bounded from above at each order in conventional perturbation. Checks of superadditivity or convexity can be carried out by retaining the terms that contribute to $\Delta''(Q)$ at the lowest possible order. Unless there is accidental cancellation or suppression at work at this order, there is no need to go to higher orders since they cannot change the convexity when conventional perturbation theory is valid.

At large values of the charge, conventional perturbation theory breaks down~\cite{Badel:2019oxl}, but one may rely on large charge expansion from an effective field theory (EFT) point of view to write the scaling dimension $\Delta(Q)$ of the lowest-lying charge $Q$ operator as\footnote{Theories with moduli spaces have a different large-charge behaviour \cite{Hellerman:2017veg}.}~\cite{Hellerman:2015nra}
\begin{align}
\Delta(Q)=A Q^{d/(d-1)}+\rm{higher\,order}\,,
\label{eq:lce}
\end{align}
where $d$ is the spacetime dimension~\footnote{In this work we restrict ourselves to $d>2$ where large charge or semiclassical expansion is known to be applicable. For the convex charge conjecture at $d=2$ we refer the reader to ref.~\cite{Aharony:2021mpc} for discussion.}. $A$ is independent of $Q$ and $\rm{higher\,order}$ represents terms suppressed by some positive powers of $Q$ in the large charge limit. Neglecting the higher order terms, the second derivative of $\Delta(Q)$ can be readily obtained
\begin{align}
\Delta''(Q)=\frac{Ad}{(d-1)^2}Q^{(2-d)/(d-1)}\,.
\end{align}
For $d>0$ and unitary theories one must have $A>0$ and thus $\Delta''(Q)>0$~\footnote{However, in ref.~\cite{Aharony:2021mpc} and this work a number of non-unitary theories are tested whose non-unitarity shows different sources of origin. In such cases one can resort to additional argument such as the one given in Appendix A of ref.~\cite{Antipin:2021jiw} in order to determine the sign of $A$.}. Convexity and superadditivity should thus hold for sufficiently large values of the charge. Now the reason why in three versions of the convex charge conjecture we require the distance index of the weight $w_0$ to be of $\mathcal{O}(1)$ is clear, as in the large charge regime such convexity property is expected to hold and the nontrivial parts of the conjectures thus come from the small charge and intermediate charge regimes.

The situation is a bit more obscured in the intermediate charge regime in which neither conventional perturbation nor large charge EFT is under good control. It turns out that semiclassical methods~\cite{Badel:2019oxl,Giombi:2020enj} based on an expansion in powers of some small coupling at fixed values of the 't Hooft coupling (corresponding to coupling times charge) provide a valid description from small to large charge regime, including the intermediate charge transition range. The semiclassical (next-to-)leading-order computation resums to all orders the terms in conventional perturbation that have (next-to-)leading powers of the charge. A generic semiclassical expansion for $\Delta(Q)$ takes the form
\begin{align}
\Delta(Q)=\frac{1}{\lambda_\ast}\Delta_{-1}(\lambda_\ast Q)+\Delta_{0}(\lambda_\ast Q)
+\lambda_\ast\Delta_{1}(\lambda_\ast Q)+...\,,
\end{align}
in which $\lambda_\ast$ denotes the fixed point coupling (or some other equivalent small parameter, such as $\epsilon$ in the $\epsilon$-expansion, and $1/N$ in the large $N$ expansion). It is worth noting that in the case of conventional perturbation theory, the leading-order contribution to $\Delta(Q)$, i.e. the classical scaling dimension, is linear in $Q$ which does not contribute to the convexity or superadditivity analysis. Thus in conventional perturbation one must go to next-to-leading order at least in order to pin down the convexity property of $\Delta(Q)$. The situation becomes different for semiclassical computations. \emph{In general, at semiclassical leading order, the knowledge of $\Delta_{-1}$ is sufficient to determine the convexity property of $\Delta(Q)$ for small values of $\lambda_\ast$.} This holds for generic functions $\Delta_{-1},\Delta_{0},...$ unless the contribution from $\Delta_{-1}$ is accidentally suppressed.
To see this more clearly, let us consider the scaling dimension $\Delta_Q$ of the $Q$-index traceless symmetric tensor in critical $O(N)$ model in $d=4-\epsilon$ dimensions. In conventional perturbation theory and up to $\mathcal{O}(\epsilon^2)$, it reads~\cite{Antipin:2020abu}
\begin{align}
  & {{\Delta }_{Q}}=\underbrace{Q}_{\text{SLO}}+\left[ \underbrace{-\left( \frac{1}{2}+\frac{1}{8+N} \right)Q}_{\text{SNLO}}+\underbrace{\frac{{{Q}^{2}}}{8+N}}_{\text{SLO}} \right]\epsilon  \nonumber \\
 & +\left[ \underbrace{\frac{184+N(14-3N)}{4{{(8+N)}^{3}}}Q}_{\text{SNNLO}}+\underbrace{\frac{(N-22)(N+6)}{2{{(8+N)}^{3}}}{{Q}^{2}}}_{\text{SNLO}}+\underbrace{\frac{2}{{{(8+N)}^{2}}}{{Q}^{3}}}_{\text{SLO}} \right]{{\epsilon }^{2}} \nonumber \\
 & +\mathcal{O}({{\epsilon }^{3}})\,.
\end{align}
Here the labels ``SLO, SNLO, SNNLO'' denote ``semiclassical leading order, semiclassical next-to-leading order, semiclassical next-to-next-to-leading order'' respectively~\footnote{$\epsilon$ and the appropriately normalized fixed point coupling differ by higher order terms and this leads to a difference at higher orders in the semiclassical expansion expressed in terms of $\epsilon$ compared to semiclassical expansion expressed in terms of the coupling. This subtlety however does not affect the argument made in this section.}. The crucial observation is that for \emph{each} order in semiclassical expansion, the term leading in conventional perturbation is linear in $Q$ and thus does not contribute to convexity analysis. For example, at semiclassical leading order, the term leading in conventional perturbation is $Q$, which does not contribute to convexity. When we go to semiclassical next-to-leading order, the term leading in conventional perturbation is $-\left( \frac{1}{2}+\frac{1}{8+N} \right)Q$ which is again linear in $Q$ and thus does not contribute to convexity. After subtracting the terms linear in $Q$, it is clear that in the remaining terms, the leading contribution to convexity still comes from the semiclassical leading order result, while the semiclassical next-to-leading order result is suppressed by an additional factor of $\epsilon$.

In practice, one needs to specify the intermediate charge regime more clearly. This can be achieved by considering the chemical potential-charge relation that is derived in each semiclassical computation at leading order. For example, for the $U(1)$-symmetric field theory of a complex scalar field $\phi$
\begin{align} \label{U(1)lagrangian}
\mathcal{L}=\partial\phi\partial\bar{\phi}+\frac{\lambda_0}{4}(\bar{\phi}\phi)^2
\end{align}
in $d=4-\epsilon$ dimensions, the chemical potential-charge relation at the Wilson-Fisher fixed point reads~\cite{Badel:2019oxl}
\begin{align} \label{U(1)FP}
R{{\mu }_{*}}=\frac{{{3}^{1/3}}+{{\Bigg[ 9\frac{{{\lambda }_{*}}Q}{{{(4\pi )}^{2}}}-\sqrt{81{{\Big[ \frac{{{\lambda }_{*}}Q}{{{(4\pi )}^{2}}} \Big]}^{2}}-3} \Bigg]}^{2/3}}}{{{3}^{2/3}}{{\Bigg[ 9\frac{{{\lambda }_{*}}Q}{{{(4\pi )}^{2}}}-\sqrt{81{{\Big[ \frac{{{\lambda }_{*}}Q}{{{(4\pi )}^{2}}} \Big]}^{2}}-3} \Bigg]}^{1/3}}}\,,
\end{align}
in which $Q$ is the fixed charge, $\lambda_*$ is the fixed point coupling, and $R$ is the cylinder radius.
The right-hand-side can be Taylor expanded around $\lambda_*=0$, resulting in
\begin{align}
R{{\mu }_{*}}=1+\frac{{{\lambda }_{*}}Q}{16{{\pi }^{2}}}-\frac{3\lambda _{*}^{2}{{Q}^{2}}}{512{{\pi }^{4}}}+\cdots\,.
\end{align}
We use as a criterion for the intermediate charge if on the right-hand-side of this expansion the second term is comparable to the first term, that is
\begin{align}
\lambda_t\equiv\frac{{{\lambda }_{*}}Q}{16{{\pi }^{2}}}\approx 1\,.
\end{align}
The small charge, intermediate charge and large charge regimes are then characterized by
$\lambda_t\ll 1,\lambda_t\sim\mathcal{O}(1),\lambda_t\gg 1$, respectively. This method of determining the intermediate charge regime can be adapted to other models in an obvious manner.

In Section~\ref{sec:toc} when we perform tests of the convex charge conjectures, we mainly rely on drawing plots of the scaling dimension (or its second derivative) computed via semiclassical methods as a function of charge. Although all plots are necessarily restricted to a finite range of charge (up to some maximal value), the generality of our analysis is in fact not limited by the charge cutoff. The reason can be understood from our discussion above: at large charge one may rely on expectation from large charge EFT to derive the convexity of the scaling dimension, while at small values of charge conventional perturbation theory can be used to reliably determine the convexity property. The genuine new information about charge convexity from semiclassical methods consists in knowledge of the scaling dimension in the intermediate charge regime, when both conventional perturbation and large charge EFT are not accurate enough. Semiclassical methods provide a way to reliably interpolate between the small and large charge regimes, and plots that cover this crucial transition regime with a full cover of small charge regime and a finite extension into the large charge regime allows us to produce conclusive statements about the convexity property of the scaling dimension for all charge values.

\section{Tests of the Conjectures: Refinements and New Results}
\label{sec:toc}

\subsection{Refining and extending the quartic $O(N)$ model in $4-\epsilon$ dimensions}
\label{sec:oNmodel_4-epsilon}

 \begin{figure}
\centering
\begin{subfigure}{.5\textwidth}
  \centering
  \includegraphics[width=.9\columnwidth]{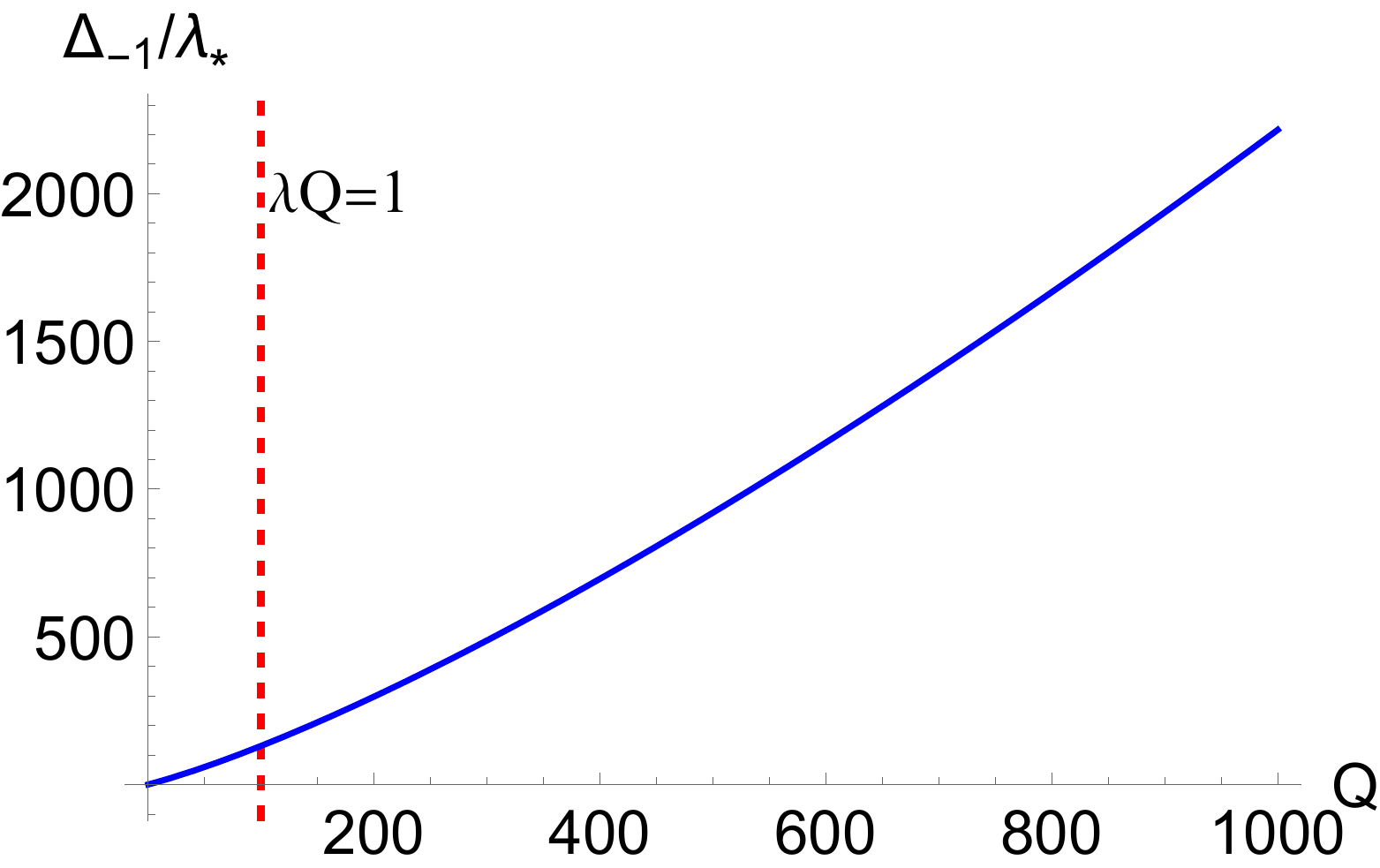}
\end{subfigure}%
\begin{subfigure}{.5\textwidth}
  \centering
  \includegraphics[width=.9\columnwidth]{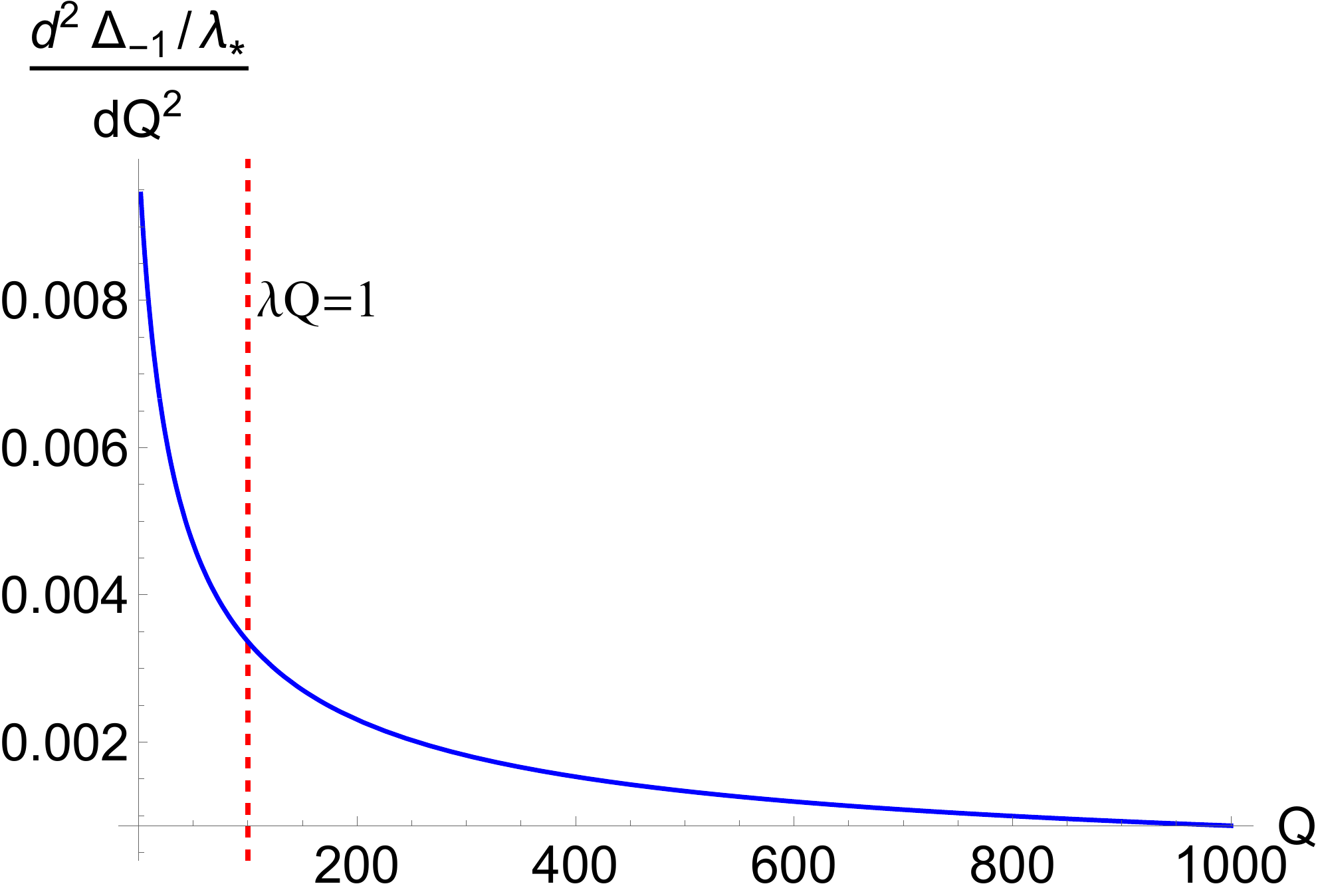}
\end{subfigure}
\caption{In the left figure, we show the semiclassical leading order scaling dimension as a function of charge $Q$ while in the right one, we show $\frac{d^2\Delta_{-1}/\lambda_*}{dQ^2}$ as a function of charge $Q$ for $O(N)$ model in $4-\epsilon$ dimension. We have chosen $\lambda_*=0.01$.}
\label{ONLeading_tot}
\end{figure}

In the work of \cite{Aharony:2021mpc}, the authors considered the $O(N)$ model at small fixed charge amenable to perturbation theory. The Lagrangian reads
 \be
{\cal L} = \frac12 \partial^{\mu} \phi_i \partial_{\mu} \phi_i + \frac{\left(4\pi\right)^2\lambda}{16} \left(\phi_i \phi_i \right)^2\;,
\ee
 with $i=1,\cdots,N$. In $d=4-\epsilon$ dimensions the fixed point is at
\be
\lambda_{\ast} = \frac{ 2\epsilon}{\left(N+8\right)} + \frac{6\left(3N+14\right)\epsilon^2}{\left(N+8\right)^3} + {\cal O}\left(\epsilon^3\right)\;.
\ee
 \begin{figure}
\centering
\begin{subfigure}{.5\textwidth}
  \centering
  \includegraphics[width=.9\columnwidth]{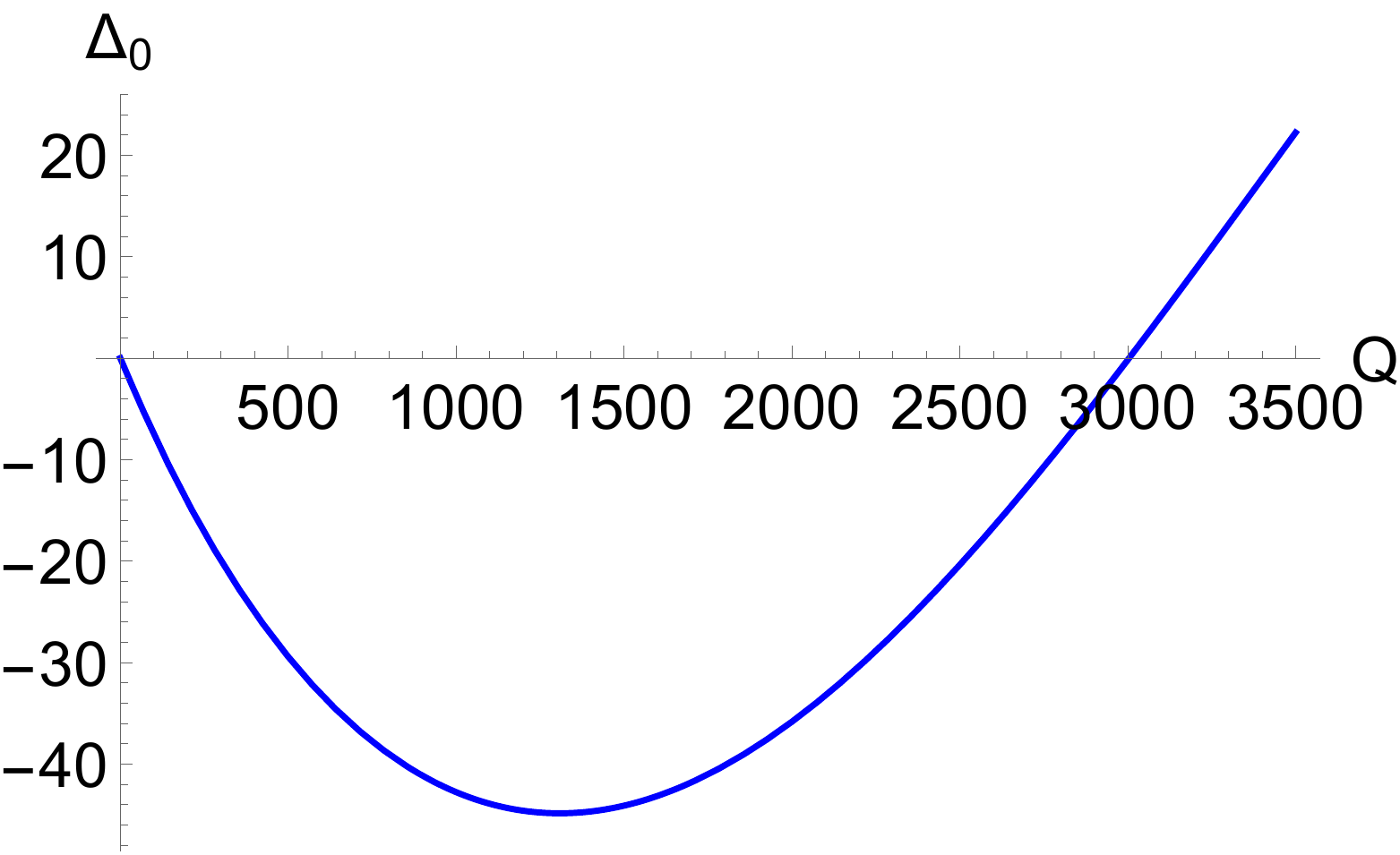}
\end{subfigure}%
\begin{subfigure}{.5\textwidth}
  \centering
  \includegraphics[width=.9\columnwidth]{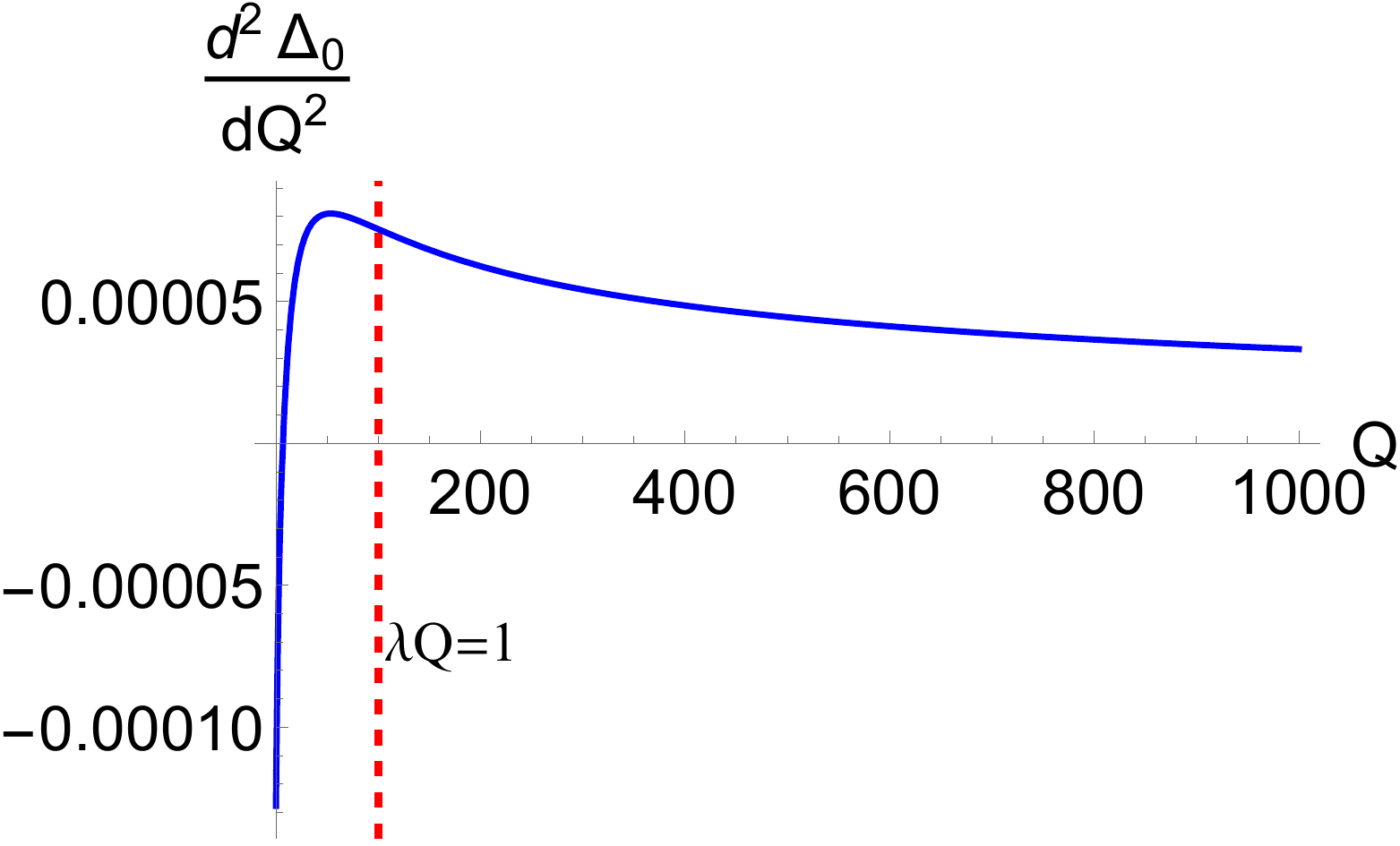}
\end{subfigure}
\caption{In the left figure, we show the semiclassical next-to-leading order scaling dimension as a function of charge $Q$ while in the right one, we show the $\frac{d^2\left(\Delta_{0}\right)}{dQ^2}$ as a function of charge $Q$ for $O(N)$ model in $4-\epsilon$ dimension. We have chosen $\lambda_*=0.01$ and $N=10$.}
\label{ONNextLeading}
\end{figure}

\begin{figure}[!b]
\centering
\begin{subfigure}{.5\textwidth}
  \centering
  \includegraphics[width=.9\columnwidth]{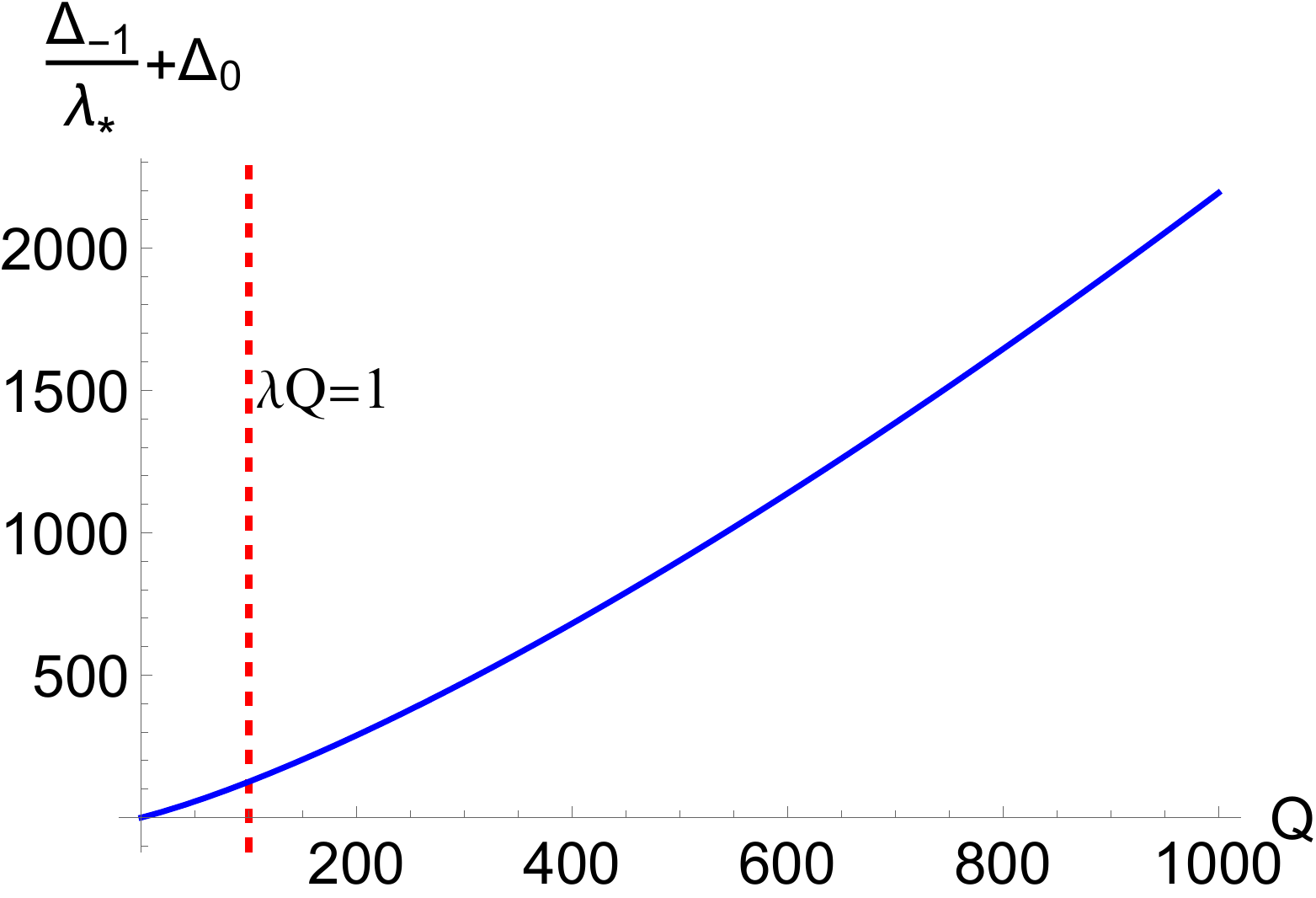}
\end{subfigure}%
\begin{subfigure}{.5\textwidth}
  \centering
  \includegraphics[width=.9\columnwidth]{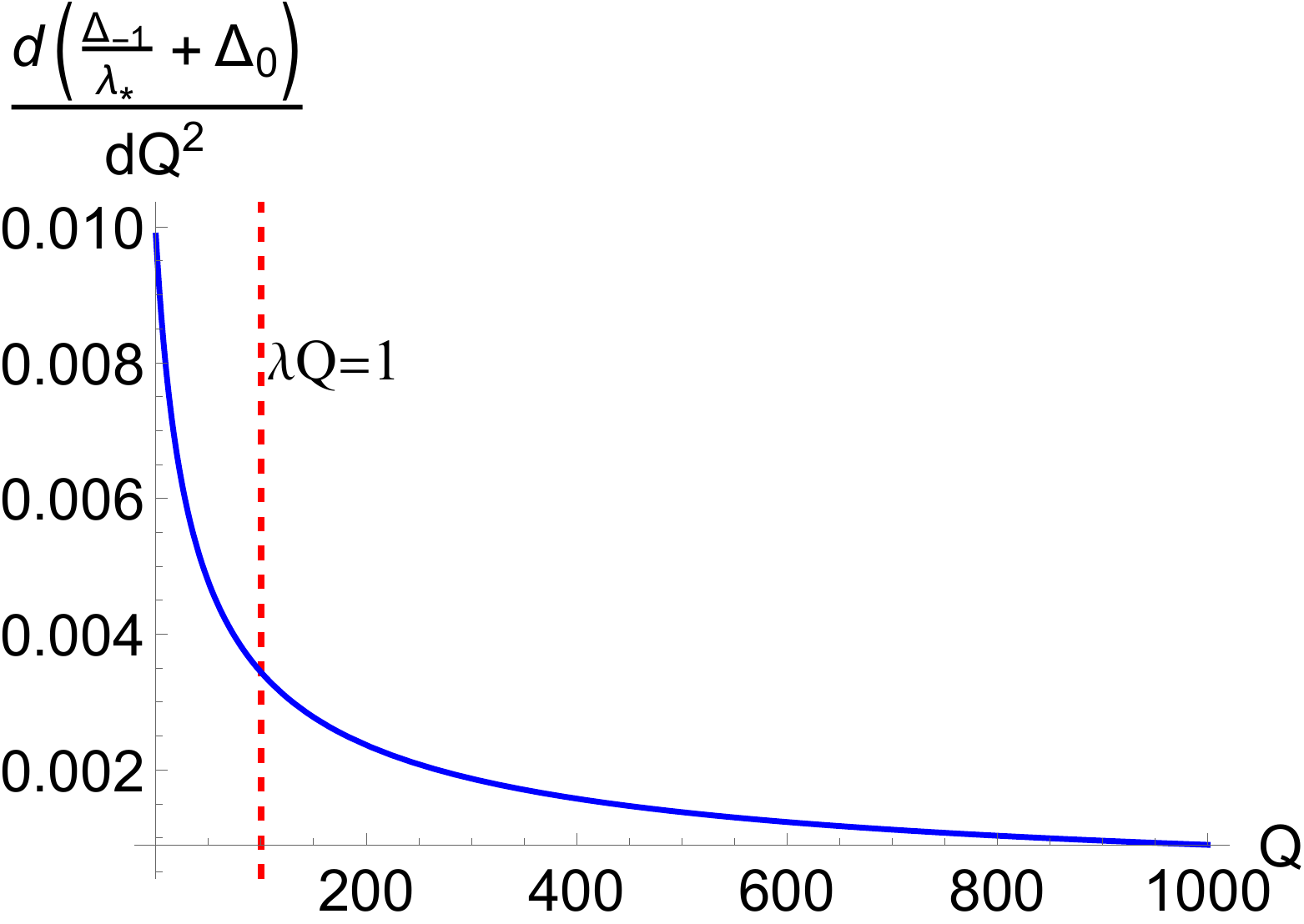}
\end{subfigure}
\caption{In the left figure, we show the total scaling dimension of semiclassical leading plus next-to-leading order as a function of charge $Q$ while in the right one, we show the $\frac{d^2\left(\frac{\Delta_{-1}}{\lambda_*}+\Delta_{0}\right)}{dQ^2}$ as a function of charge $Q$ for $O(N)$ model in $4-\epsilon$ dimension. We have chosen $\lambda_*=0.01$ and $N=10$.}
\label{ONLeading+NLL_tot}
\end{figure}

We consider the class of operators $\varphi^Q$, where $\varphi \equiv \phi_1  + i \phi_2$, which live in the $Q$-index traceless symmetric tensor representation of $O(N)$. For small charge $Q$ and based on conventional perturbative results at one loop order, the authors of \cite{Aharony:2021mpc} have shown that the inequality Eq.~\eqref{key_equation} is satisfied.

 \begin{figure}[!t]
\centering
	\includegraphics[width=0.7\textwidth]{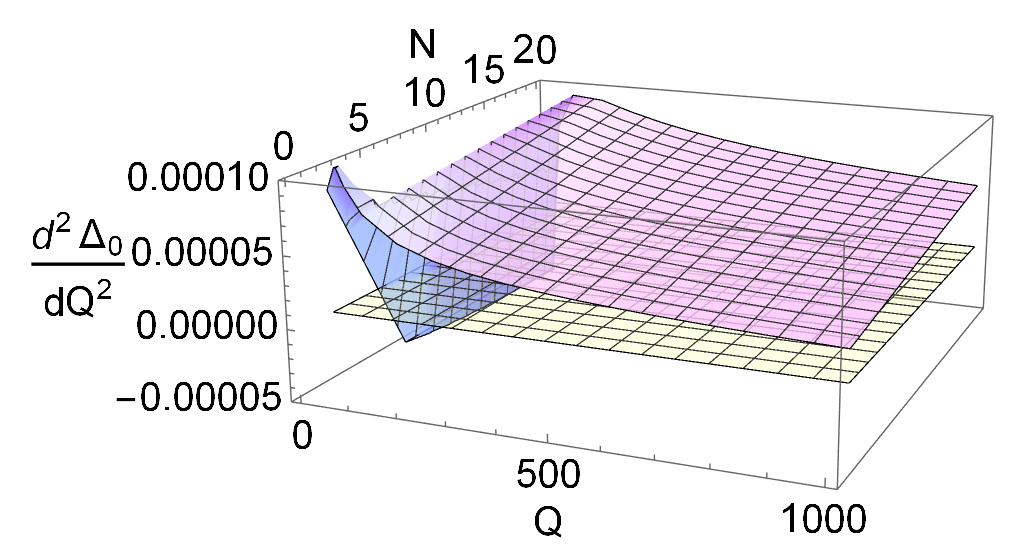}
	\caption{In this figure, we show $\frac{d^2\Delta_{0}}{dQ^2}$ (purple) as a function of charge $Q$ and variable $N$ for $O(N)$ model in $4-\epsilon$ dimension. We have chosen $\lambda_*=0.01$. The yellow sheet is chosen to be at $0$ as a reference.}
	\label{N-dependenceofD0_derivative}
\end{figure}

Here we extend this result to fixed 't Hooft coupling $ \lambda_{*} Q $ while expanding in $ \lambda_{*}$ within the semiclassical framework:
\begin{equation} \label{expform}
\Delta_{Q}=\frac{1}{\lambda_*}\Delta_{-1}\left({\lambda_*\,Q}\right)+\Delta_{0}\left({\lambda_*\,Q}\right)+\lambda_*\Delta_{1}\left({\lambda_*\,Q}\right)+\cdots\,.
\end{equation}
According to the current state-of-the-art  we have the full knowledge of the $\Delta_{-1}\left({\lambda_*\,Q}\right)$  and $\Delta_{0}\left({\lambda_*\,Q}\right)$ functions. These are given in \cite{Antipin:2020abu}. It is therefore instructive to investigate independently their structure and, in particular, whether they are independent convex functions or only the sum is convex.

We start by reporting the analytical expression for $\Delta_{-1} $ that reads:
 \begin{equation}
\Delta_{-1}\left({\lambda_*\,Q}\right)=\frac{1}{32} \left(3 b^4-2 b^2-1\right)\,,
\label{ON_LL_Semi}
\end{equation}
 where
\begin{equation}
b\equiv\frac{\left(\sqrt{1296 J^2-3}+36 J\right)^{2/3}+\sqrt[3]{3}}{3^{2/3} \sqrt[3]{\sqrt{1296 J^2-3}+36 J}},\,\qquad J=\frac{\lambda_\ast\,Q}{4}\,.
\label{ON_LL_Semi_chemical}
\end{equation}
The convexity of this function is clear from the plots of the function itself  given in the left \autoref{ONLeading_tot}  and from the positivity of its second derivative $\frac{d^2\Delta_{-1}/\lambda_*}{dQ^2}$ shown in right \autoref{ONLeading_tot}.

This result is  interesting since one could argue towards a positive test  of the \cite{Aharony:2021mpc} convexity conjecture  at  intermediate charges.   Nevertheless it is instructive to also analyze the convexity properties of  $\Delta_0$. Differently from $\Delta_{-1}$ we  have a numerical result for $\Delta_0$  that we report in \autoref{ONNextLeading}. From it we observe that $\Delta_0$ is convex except for very small 't Hooft coupling where $\lambda_* Q\leq 0.1$. Therefore, we expect the sum of leading and next-to-leading order scaling dimension is also convex as shown in the right of \autoref{ONLeading+NLL_tot} where we plot also the second derivative of $\frac{\Delta_{-1}}{\lambda_*} +\Delta_{0}$ which is clearly positive.

 \begin{figure}[!t]
\centering
	\includegraphics[width=0.7\textwidth]{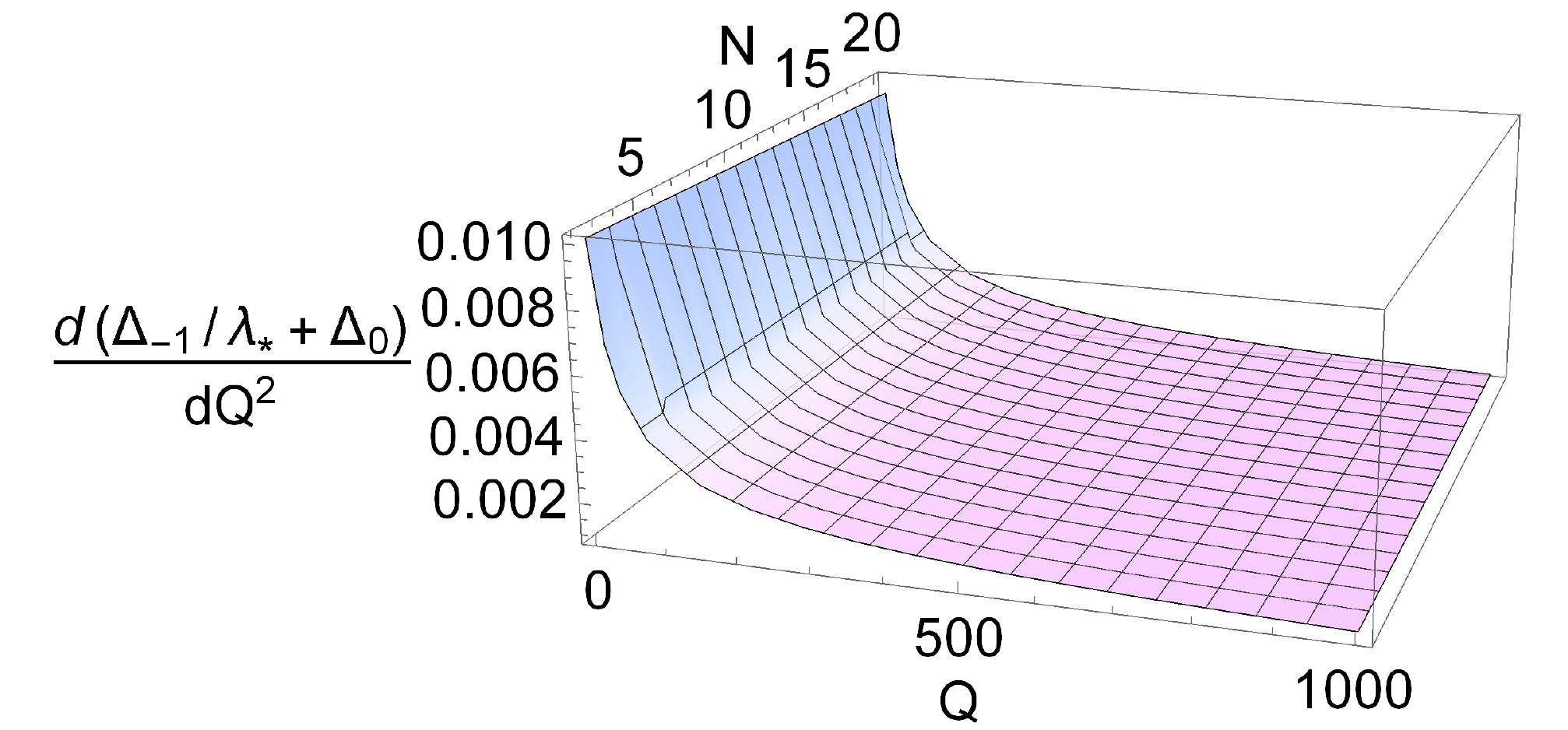}
	\caption{In this figure, we show $\frac{d^2\left(\Delta_{-1}/\lambda_{*}+\Delta_{0}\right)}{dQ^2}$ as a function of charge $Q$ and variable $N$ for $O(N)$ model in $4-\epsilon$ dimension. Without losing generality, we have chosen $\lambda_*=0.01$.}
	\label{3DO(N)LL+NLL_Curvature}
\end{figure}

In fact, the $\Delta_{-1}$ function is convex for any charge $Q$ and variable $N$ while we have checked that $\Delta_{0}$ is convex up to $Q\sim 10^3$ and $N=5\--20$ except for the region of very small 't Hooft coupling ($\lambda Q\leq 0.1$) showing the concave behaviour. It is reasonable to expect a similar feature to hold also for $N>20$, and can be further numerically checked later. For $N=2,\,3,\,4$, however, $\Delta_{0}$ is convex up to $Q\sim 10^3$ since the concave region actually occurs at $Q<1$. These features are reported in \autoref{N-dependenceofD0_derivative} where we show respectively $\Delta_{0}$ and $\frac{d^2\Delta_0}{dQ^2}$ for $N=2\--20$ and $Q=0\--1000$.  Note that because the $N$ dependence of $\lambda_\ast$ can be traded for a different value of $\epsilon$ we are entitled to keep it constant. Finally, in \autoref{3DO(N)LL+NLL_Curvature}, we show the convexity property of the sum of the leading and next-to-leading results i.e.~$\frac{d^2\left(\Delta_{-1}/\lambda_{*}+\Delta_{0}\right)}{dQ^2}$ as a function of charge $Q$ and variable $N$. For all the plots here we chose $\lambda_\ast = 0.01$.  Changing the value of $\lambda_\ast$ corresponds to a modified range of $Q$ which does not affect the results. However, it should be noted that if we choose a bigger coupling value, say $\lambda_*\sim0.1$, then the small concave region in $\Delta_{0}$ disappears since it requires charge $Q\leq 1$. The $U(1)$  case is trivially included in the analysis for $N=2$.

We have therefore shown that the conjecture holds also in the intermediate charge regime and for different intermediate values of $N$. Interestingly, we discover that the leading and the sum of leading and subleading terms in the semiclassical framework abide the conjecture. The subleading term is convex except for the region of very small 't Hooft coupling.

\subsection{Refining and extending the quartic $O(N)$ model in $4+\epsilon$ dimensions}
\label{subsec:3b}

\begin{figure}[!t]
\centering
\begin{subfigure}{.5\textwidth}
  \centering
  \includegraphics[width=.9\columnwidth]{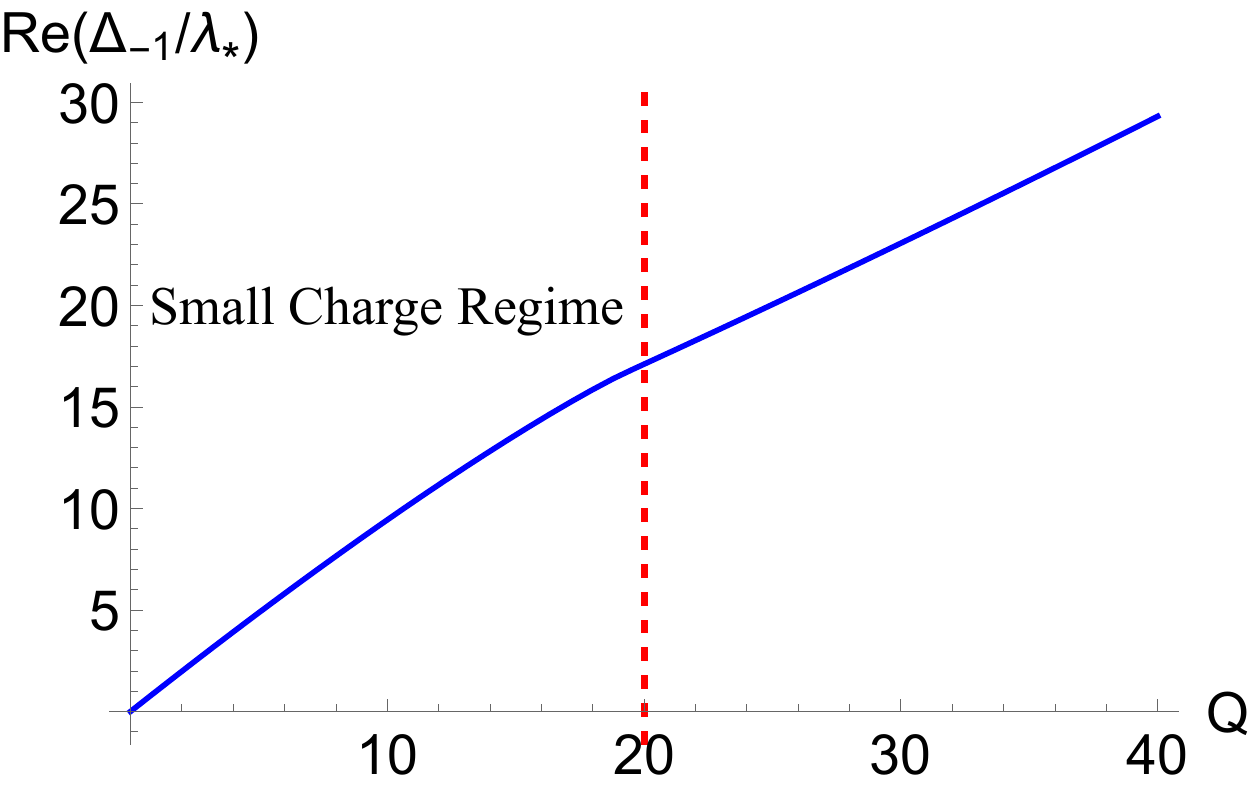}
\end{subfigure}%
\begin{subfigure}{.5\textwidth}
  \centering
  \includegraphics[width=.9\columnwidth]{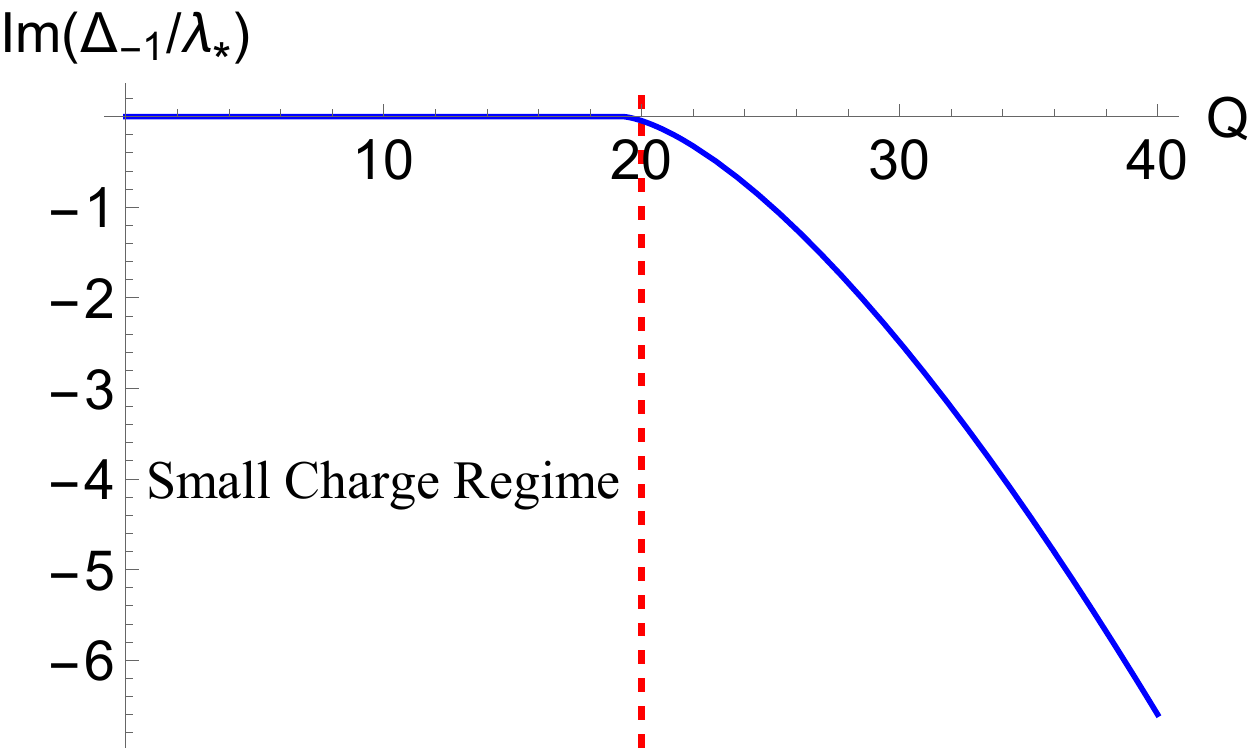}
\end{subfigure}
\caption{We show the real and imaginary part of the semiclassical leading order scaling dimension $\Delta_{-1}/\lambda_*$ as a function of charge $Q$ in $4+\epsilon$ dimension for $O(N)$ quartic model in respectively the left and right part of the figure. In each sub-figure, the left side of the red dashed line represents the small charge regime with $\lambda_* Q=-0.2$. We have chosen $\lambda_*=-0.01$.}
\label{ONLeading_4+epsilon}
\end{figure}
In this section, we consider $O(N)$ quartic model in $4+\epsilon$ dimensions which features an ultraviolet fixed point at negative coupling values . We can directly apply the leading-order semiclassical results Eq.~\eqref{ON_LL_Semi} and Eq.~\eqref{ON_LL_Semi_chemical} in the above section \autoref{sec:oNmodel_4-epsilon}. This case is particularly interesting since in the small charge regime the scaling dimension is real while in the large charge regime it becomes complex. In fact, as pointed out in \cite{Antipin:2021jiw, Giombi:2020enj}, the model exhibits a critical value of the charge where $\Delta_{-1}$ develops a branch cut and above which the scaling dimensions acquire a nonzero imaginary part. To the leading order in the $\epsilon$-expansion, the critical value of the charge reads \cite{Antipin:2021jiw}
\begin{equation}
    Q_c  = \frac{N+8}{6 \sqrt{3} \epsilon}\,.
\label{branch_cut}
\end{equation}
In the case of complex scaling dimension it is the real part of the scaling dimension that inherits the role of the energy of the corresponding state when we perform a Weyl map to the cylinder, and thus it is well-motivated to examine the convexity property of the real part of the scaling dimension.
At small values of the charge, the scaling dimension of $\varphi^Q$ operator can be obtained by making the replacement $\epsilon\rightarrow -\epsilon$ in the corresponding expression for the $O(N)$ model in $4-\epsilon$ dimensions, from which we may conclude that the real part of the scaling dimension is concave with respect to the charge. For large values of the charge,
as long as the prefactor $A$ in Eq.~\eqref{eq:lce} has a positive real part~\footnote{One can prove the positivity of the real part of $A$ using the argument given in Appendix A of ref.~\cite{Antipin:2021jiw}.}, we expect the real part of the scaling dimension to be convex with respect to the charge. Therefore, a transition from concave to convex behavior for the real part of the scaling dimension as a function of the charge is expected, if the above picture is correct.

\begin{figure}
\centering
	\includegraphics[width=0.6\textwidth]{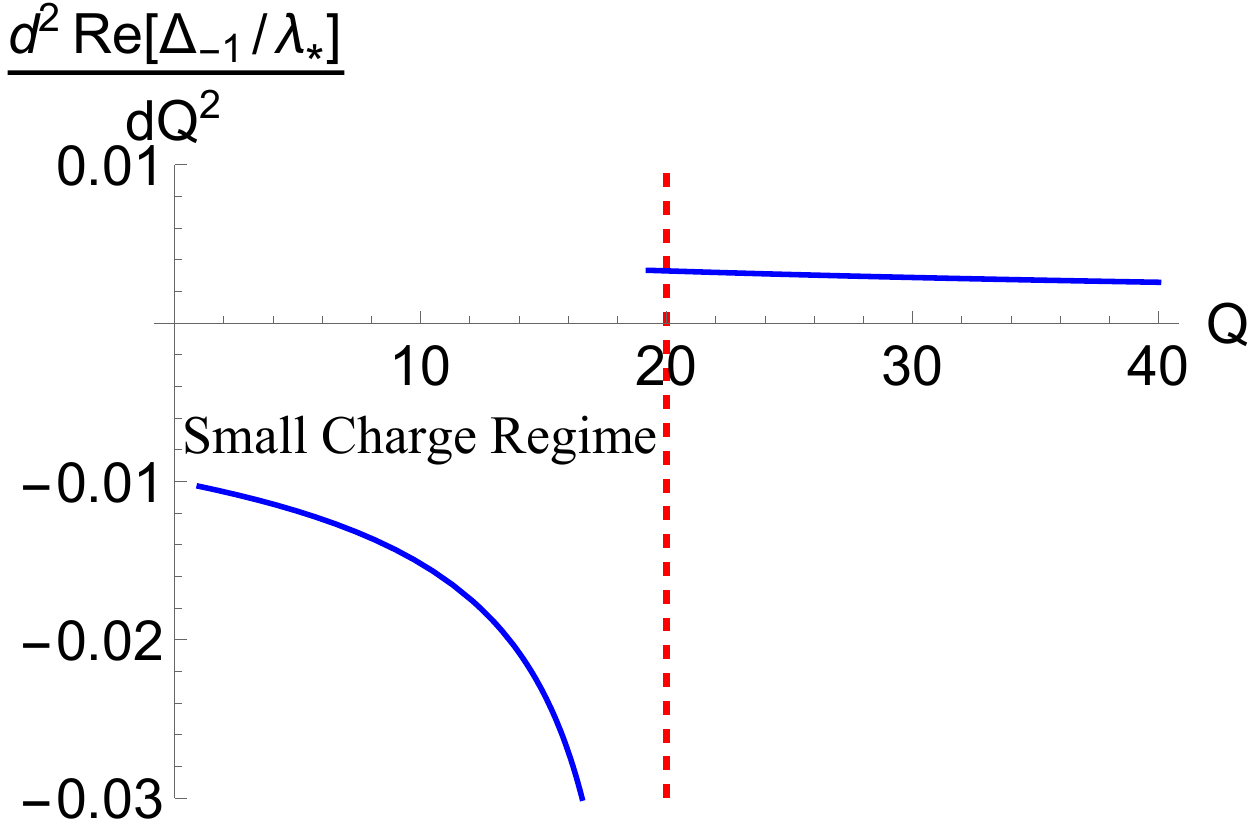}
	\caption{In this figure, we show the convexity property of the real part of the semiclassical leading-order scaling dimension $\frac{d^2\Delta_{-1}/\lambda_*}{dQ^2}$ as a function of charge $Q$. The left side of the red dashed line represents the small charge regime with $\lambda_* Q=-0.2$. Without losing generality, we have chosen $\lambda_*=-0.01$.
	}
	\label{ONLeading_4+epsilon_Real_Curvature}
\end{figure}

We present the real and imaginary part of the leading order scaling dimension $\Delta_{-1}/\lambda_*$ as a function of charge $Q$ in $4+\epsilon$ dimension in \autoref{ONLeading_4+epsilon}. In each sub-figure, we also include a reference line (shown as a red dashed line) which provides a reference 't Hooft coupling value $\lambda_* Q=-0.2$. Here, we have chosen $\lambda_*=-0.01$. With the parameters we have chosen, using Eq.~\eqref{branch_cut}, we can determine $Q_c\sim 20$. It is clear from the left part of \autoref{ONLeading_4+epsilon} that the real part of the scaling dimension is concave below $Q_C$ while becomes convex above it. This is further shown in \autoref{ONLeading_4+epsilon_Real_Curvature}. In the right part of \autoref{ONLeading_4+epsilon}, we observe that the imaginary part of the scaling dimension is zero below $Q_C$ and thus the scaling dimension is indeed real in the small charge regime.

\subsection{Refining and extending the sextic $O(N)$ model in $3-\epsilon$ dimensions}
\label{sec:u1model3e}

\begin{figure}
\centering
\begin{subfigure}{.5\textwidth}
  \centering
  \includegraphics[width=.9\columnwidth]{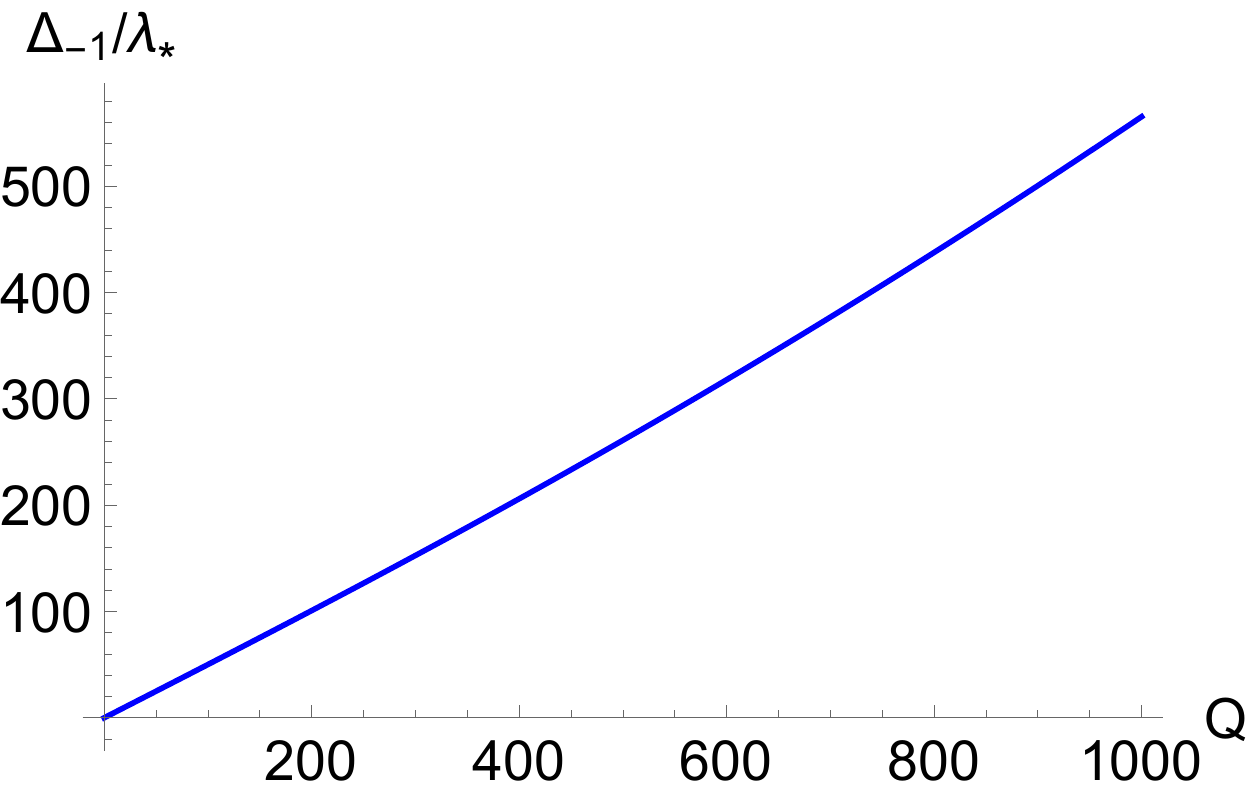}
\end{subfigure}%
\begin{subfigure}{.5\textwidth}
  \centering
  \includegraphics[width=.9\columnwidth]{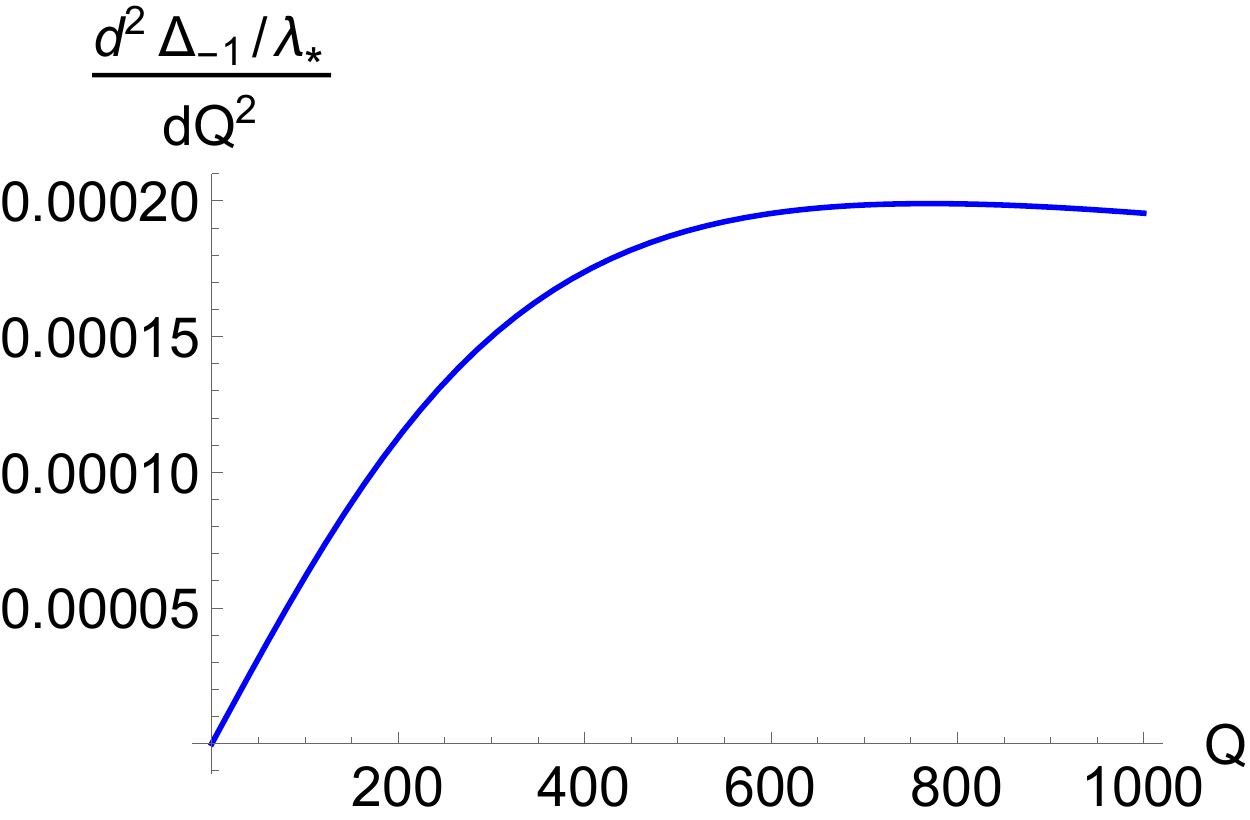}
\end{subfigure}
\caption{In the left figure, we show the semiclassical leading order scaling dimension as a function of charge $Q$ while in the right one, we show the $\frac{d^2\Delta_{-1}/\lambda_*}{dQ^2}$ as a function of charge $Q$. Both figures are discussed within $O(N)$ model in $3-\epsilon$ dimesnions. We have chosen $\lambda_*=0.01$.}
\label{3-epsilon_O(N)_Leading_tot}
\end{figure}

For the sextic $O(N)$ model in $d=3-\epsilon$ (where both the $|\phi|^2$ and the $|\phi|^4$ operators are fine-tuned to zero), the Euclidean Lagrangian is given by
\be
{\cal L} = \frac{1}{2}\partial^{\mu} \phi_i \partial_{\mu} \phi_i + \left(\frac{\lambda^2}{48}\right) \left(\phi_i \phi_i \right)^3 \;.
\ee

Similar to the discussions in \autoref{sec:oNmodel_4-epsilon}, here we again study the convexity conjecture Eq.\eqref{key_equation} by using the semiclassical computations. Again, we have the knowledge of $\Delta_{-1}$ and $\Delta_0$~\cite{Badel:2019khk,Jack:2020wvs}. Below we only report the convexity study using the leading-order semiclassical results, which are already sufficient to prove the convexity property. The semiclassical leading-order scaling dimension is given by:
\begin{equation}
    \frac{\Delta_{-1}}{\lambda} = Q \frac{1+ \sqrt{1+x}+x/3}{\sqrt{2}\left(1+ \sqrt{1+x} \right)^{3/2}},\,\qquad x = \frac{\lambda^2 Q^2}{2 \pi^2}\,.
\label{leading_3-epsilon}
\end{equation}
By using Eq.~\eqref{leading_3-epsilon}, in \autoref{3-epsilon_O(N)_Leading_tot}, we present the semiclassical leading order scaling dimension $\Delta_{-1}$ and its convexity property $\frac{d^2\Delta_{-1}}{dQ^2}$ as a function of charge $Q$. We have chosen $\lambda_*=0.01$. It is clear from the figure that the convexity property holds in this model.

\subsection{Refining and extending the quartic $O(N)$ model in $d$ dimensions at large $N$ test}
\label{sec:onmodelgend}

In this section we study the scaling dimensions of the quartic $O(N)$ theory by using the $1/N$ expansion \cite{Alvarez-Gaume:2019biu,Giombi:2020enj} instead of the $\epsilon$ expansion procedure used in the previous section \ref{sec:oNmodel_4-epsilon}. The $1/N$ expansion has the advantage that it applies to generic dimension $d$ and thus is able to provide a larger landscape to test the weak-gravity conjecture. Cases of $d=3$, $d=5$ and $d=6-\epsilon$ have been studied in \cite{Aharony:2021mpc} using the small $Q/N$ expansion. In this work, we directly apply the semiclassical large $N$ results shown in \cite{Giombi:2020enj} to test the weak gravity conjecture and to go beyond the small charge regime.

\begin{figure}[!b]
\centering
	\includegraphics[width=0.6\columnwidth]{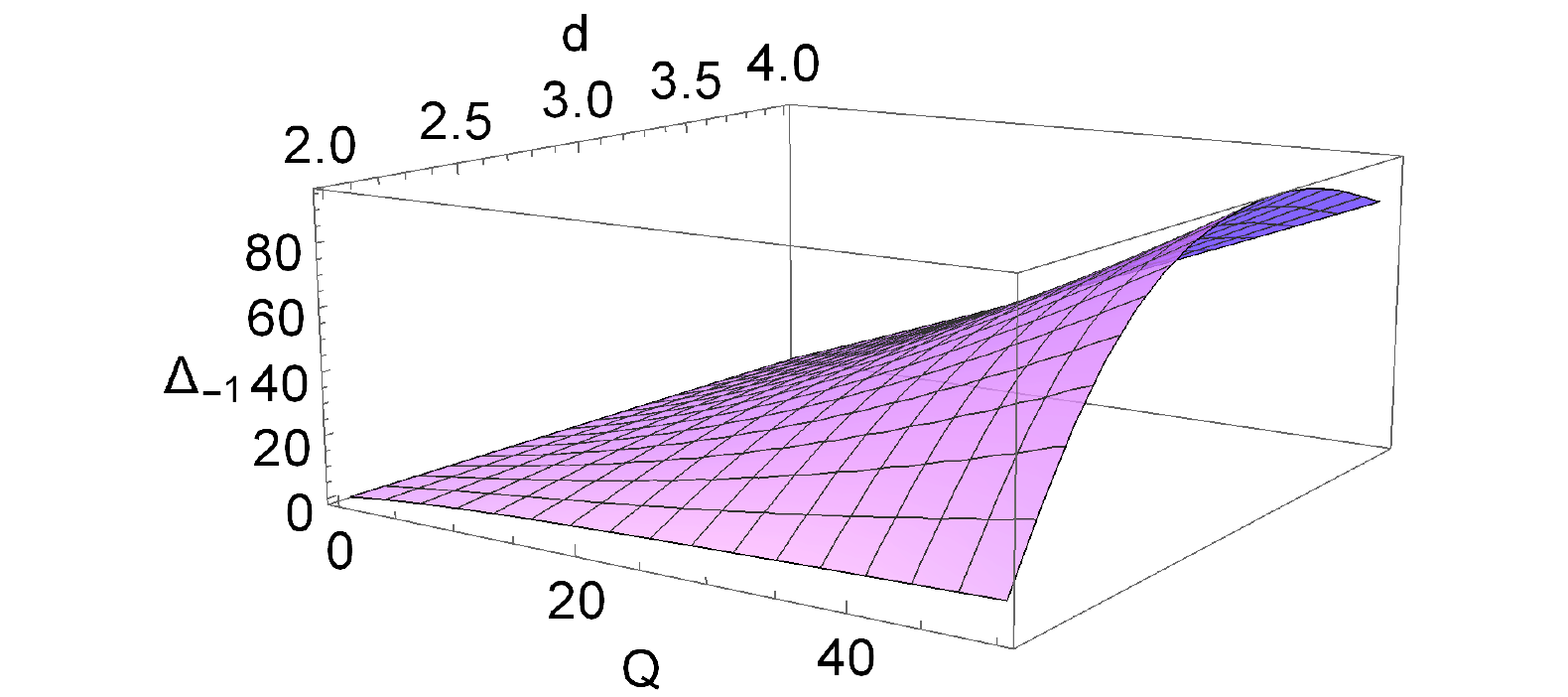}
	\caption{In this figure, we show the $O(N)$ model leading order scaling dimension $\Delta_{-1}$ at large $N$ as a function of charge $Q$ and dimension $2<d<4$. We have chosen $N=20$.
	}
	\label{LargeN_ON_Leading_d2-4}
\end{figure}

Following \cite{Giombi:2020enj}, the scaling dimension at the leading order in the semiclassical 1/N expansion can be written as :
\begin{equation}
\Delta_{-1}=N\left(f\left(c_\sigma\right)+\frac{Q}{N}\sqrt{\left(\frac{d}{2}-1\right)^2+c_\sigma}\right),\,\qquad f\left(c_\sigma\right)\equiv -\frac{c_\sigma}{d-2}\int_{0}^\infty dt \frac{J_2\left(\sqrt{c_\sigma}t\right)}{t\left(2cosht-2\right)^{\frac{d}{2}-1}}
\label{large_N}
\end{equation}
where $J_2$ is the conventional Bessel function and $c_\sigma$ is determined from:
\begin{equation}
\frac{d}{d\sigma}\left(f\left(c_\sigma\right)+\frac{Q}{N}\sqrt{\left(\frac{d}{2}-1\right)^2+c_\sigma}\right)=0\,.
\label{large_N_determine_c}
\end{equation}

\begin{figure}
\centering
	\includegraphics[width=0.7\columnwidth]{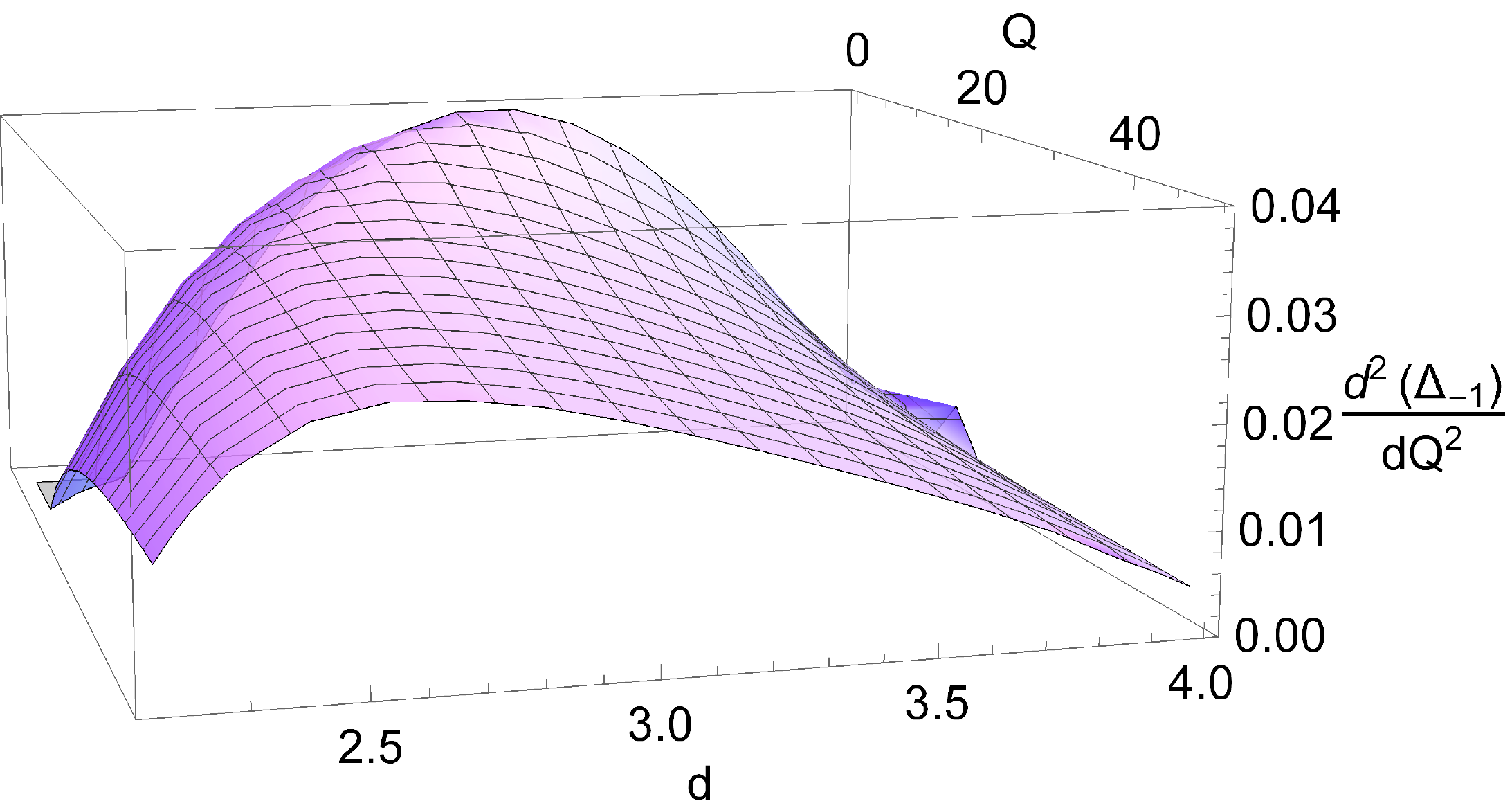}
	\caption{In this figure, we show the convexity property of the $O(N)$ model leading order scaling dimension $\frac{d^2\Delta_{-1}}{dQ^2}$ at large $N$ as a function of charge $Q$ and dimension $2<d<4$. Without losing generality, we have chosen $N=20$.
	}
	\label{LargeN_ON_Leading_d2-4_curvature}
\end{figure}

By using Eq.~\eqref{large_N} and Eq.~\eqref{large_N_determine_c}, in \autoref{LargeN_ON_Leading_d2-4} we present the $O(N)$ model leading order scaling dimension $\Delta_{-1}$  at large $N$ as a function of charge $Q$ and dimension $2<d<4$. Without losing generality, we have chosen $N=20$. To prove the convexity conjecture, in \autoref{LargeN_ON_Leading_d2-4_curvature}, we show the convexity property of the $O(N)$ model leading order scaling dimension $\frac{d^2\Delta_{-1}}{dQ^2}$ at large $N$ as a function of charge $Q$ and dimension $2<d<4$ with $N=20$. It is clear that the convexity property of the quartic $O(N)$ model holds for any dimension $2<d<4$.

As pointed out in \cite{Aharony:2021mpc}, in $4<d<6$, the large $N$ $O(N)$ model violates the conjecture at small $Q/N$ where CFT data are real. In this range of dimensions the model features a critical value of the charge $Q_c$ above which the scaling dimensions acquire a nonzero imaginary part. Interestingly, as illustrated by the model in $d=4+\epsilon$ dimensions, at $Q_c$ the convexity properties of the spectrum change in such a way that if one considers the real part of the scaling dimensions, the large-charge (large $Q/N$) regime is convex.

\subsection{New tests: The $U(N)\times U(M)$ in $4-\epsilon$ dimensions}
\label{subsec:unumtest}

We consider the $U(N)\times U(M)$ linear sigma model in $d=4-\epsilon$
\begin{align}
\mathcal{L}=\Tr(\del_\mu H^\dagger \del^\mu H ) + u_0\Tr(H^\dagger H)^2 + v_0(\Tr H^\dagger H )^2\,.
\end{align}
Here $H$ is an $N\times M$ complex matrix scalar field. Without loss of generality we consider $N\leq M$. The global symmetry of the model is
\begin{equation}
\mathcal{G} \equiv SU(N)_L\times SU(M)_R\times U(1)_A\,,
\end{equation}
where $U(1)_A$ is an universal phase rotation of $H$. The Noether charges are encoded in a traceless matrix defined as
\begin{equation}
    \mathcal{Q} = \frac{i}{2} \int d^3 x \left(\dot H H^\dagger - H \dot H^\dagger \right)\,.
\end{equation}

\begin{figure}[!t]
\centering
	\includegraphics[width=0.65\columnwidth]{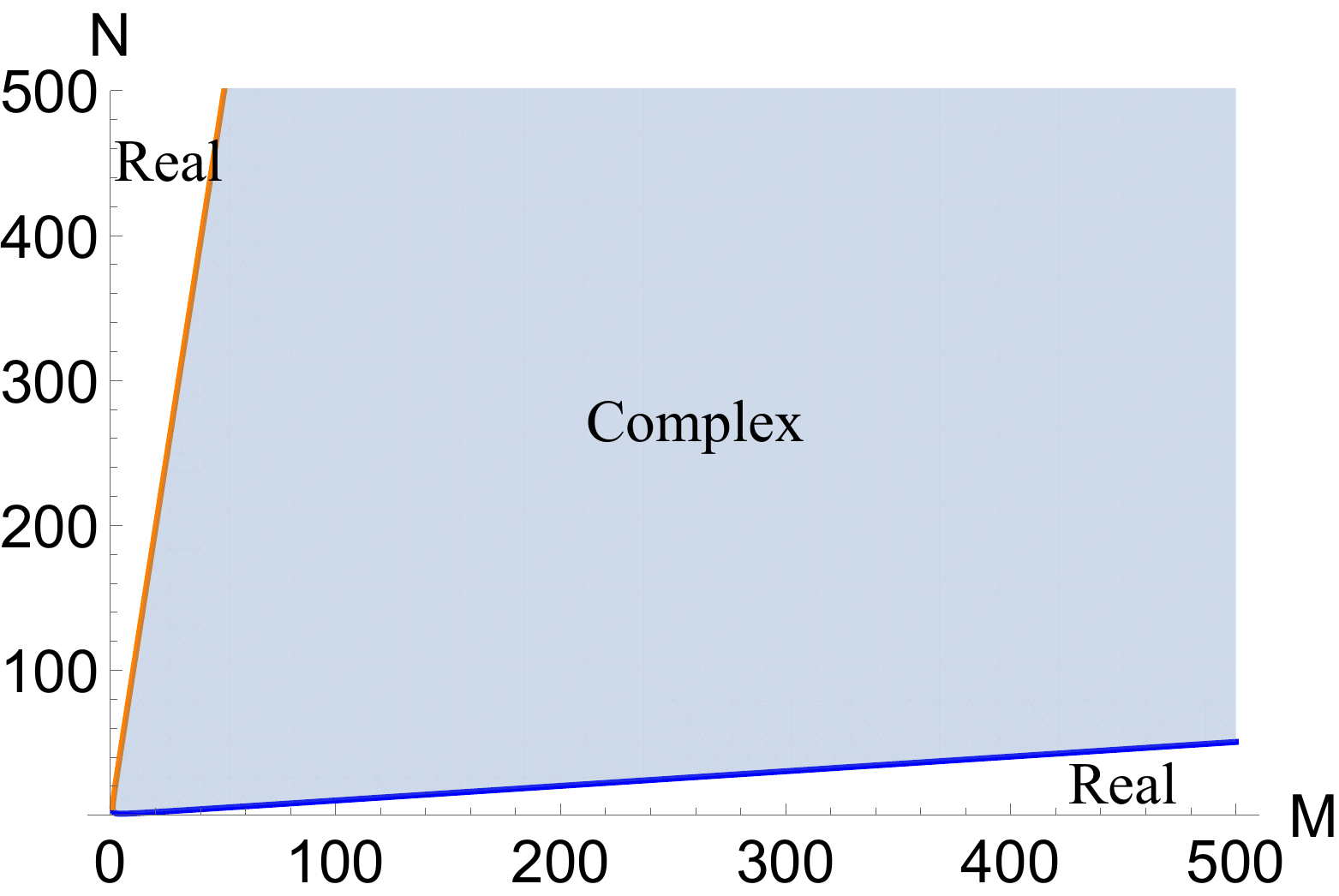}
	\caption{In this figure, we show the phase diagram of the $U(N)\times U(M)$ model where the fixed points are either real or complex.
	}
	\label{real_complex_phase_diagram}
\end{figure}

At the $1$-loop level the model features four FPs: a Gaussian FP ($u^* = v^* = 0$), an $O(2 N M )$ FP ($u^* = 0$) and other two FPs given by
\begin{equation} \label{complFP}
   u^*_{\pm} = 4 \pi^2 \frac{A_{M N} \mp 3 \sqrt{R_{M N}}}{ D_{M N}} \epsilon \,, \qquad \qquad v^*_{\pm} = 4 \pi^2 \frac{B_{M N} \pm (M +N) \sqrt{R_{M N}}}{2 D_{M N}} \epsilon \,,
\end{equation}
with
\begin{eqnarray}
&& A_{MN} = N M^2 + M N^2 - 5 N - 5M \,,\qquad \qquad B_{M N} = 36 - (M + N)^2 \,, \nonumber \\
&& R_{MN} = 24 + M^2 + N^2 - 10 M N \,, \qquad \qquad D_{M N} =( M N - 8) (M + N)^2 + 108 \,. \nonumber \\
\end{eqnarray}
This model displays an interesting phase structure since we can go from real to complex conformal dimensions by changing the number of matter fields $M,\,N$. In particular, the last two FP are complex when $R_{MN}<0$. In \autoref{real_complex_phase_diagram}, we show the phase diagram of the $U(N)\times U(M)$ model where the last two fixed points are either real or complex.

Below we consider two types of charge configurations. For the first charge configuration, we first study the inequality Eq.~\eqref{key_equation} by using the one-loop scaling dimensions. Then, we apply the full semiclassical results (leading and next-to-leading order) to study the convexity property.  For the second charge configuration, due to its numerical complications, we only study the inequality Eq.~\eqref{key_equation} at small $\epsilon$ by considering the one-loop scaling dimension. For more details of the semiclassical computations of $U(N)\times U(M)$ model, we refer the readers to \cite{Antipin:2020rdw,Antipin:2021akb}.

\textbf{Charge Configuration one:}\\
Consider the following $2$-parameters family of charge configurations
\begin{equation} \label{complete}
    \mathcal{Q}_{J,s} = \text{diag}\big(\underbrace{J,J, \dots}_s , \underbrace{-J, -J, \dots}_s , \underbrace{0, 0, \dots}_{N-2s} \big) \,.
\end{equation}
The semiclassical expansion takes the following form
\begin{equation} \label{forma}
    \Delta_{J}  =  \sum_{k=-1} \frac{1}{J^k} \Delta_k (\mathcal{A}_h^*, \mathcal{A}_v^*) \,.
\end{equation}
where we introduced 't Hooft-like couplings as
\begin{align}\label{eq:couplings}
\cA_h = J \frac{u N}{(4\pi)^2} \ , \qquad \cA_v = J \frac{ 2 s v N}{(4\pi)^2} \ ,
\end{align}
The leading order of the semiclassical expansion reads
\begin{equation}
    \Delta_{-1} (\mathcal{A}_h^*, \mathcal{A}_v^*) = \frac{s N}{144(\mathcal{A}_h^* + \mathcal{A}_V^* )} \frac{1}{x^{*\frac{4}{3}}}\left(\sqrt[3]{3} x^{*8/3}-3 x^{*4/3}+6 \sqrt[3]{3} x^{*2/3}+2\ 3^{2/3}
   x^{*2}+ 3^{5/3} \right) \,.
\label{leading_UN_UM}
\end{equation}
with
\begin{equation}
     x=  \frac{72}{N}(\mathcal{A}_h + \mathcal{A}_v)  + \sqrt{-3+ \left(\frac{72}{N}(\mathcal{A}_h + \mathcal{A}_v) \right)^2 } \,,
\end{equation}
The NLO can be written in terms of a convergent sum as
\begin{align} \label{NLO}
&\Delta_{0}(\cA_h^*, \cA_v^*) = \rho (x^*,M, N, s, \cA_h^*,  \cA_v^*)  + \nonumber \\
&\frac{1}{2} \sum_{\ell=0}^\infty \left[ R (1+ \ell)^2
\left(\sum_i g_i(M,N,s) \omega_i(\ell, x^*, \cA_h^*, \cA_v^* )\right)_{d=4} + \sigma(\ell, x^*, M, N, s,  \cA_h^*,  \cA_v^*) \right] \,.
\end{align}
where the explicit form of the two functions $\rho (x^*,M, N, s, \cA_h^*,  \cA_v^*)$ and $\sigma(\ell, x^*, M, N, s,  \cA_h^*,  \cA_v^*)$ can be found in \cite{Antipin:2021akb}.
From the above results, one can extract the one-loop scaling dimension of the lowest-lying operators carrying this charge configuration, which reads
\begin{align} \label{one-loop}
\Delta_{{Q=4 s J}}^{1-loop} & =Q \left(1- \frac{\epsilon}{2}\right) + \frac{4}{N} \left(A_h^* \left(Q-2 s^2\right)+(Q-1) A_v^*\right)  \nonumber \\
& = Q \left(1- \frac{\epsilon}{2}\right) + \frac{Q \left(Q-2 s^2\right)}{ (4 \pi)^2 s} u^* +\frac{2 Q (Q- 1)}{(4 \pi)^2} v^* \,,
\end{align}
where $Q =4 s J $ is, in the perturbative regime, the classical scaling dimension of the corresponding fixed-charge operator. By defining $\gamma_{Q_1,Q_2} \equiv \Delta_{Q_1+Q_2}-\Delta_{Q_1}-\Delta_{Q_2}$, such that the conjecture is satisfied when $\gamma_{Q_1,Q_2}>0$, it follows that
\begin{equation}
    \gamma_{Q_1,Q_2} =\frac{Q_1 Q_2}{8 \pi^2 s} (u^*+ 2 s v^*) + \cO\left(\epsilon^2\right)\,.
\end{equation}

\begin{figure}[!t]
\centering
	\includegraphics[width=0.85\columnwidth]{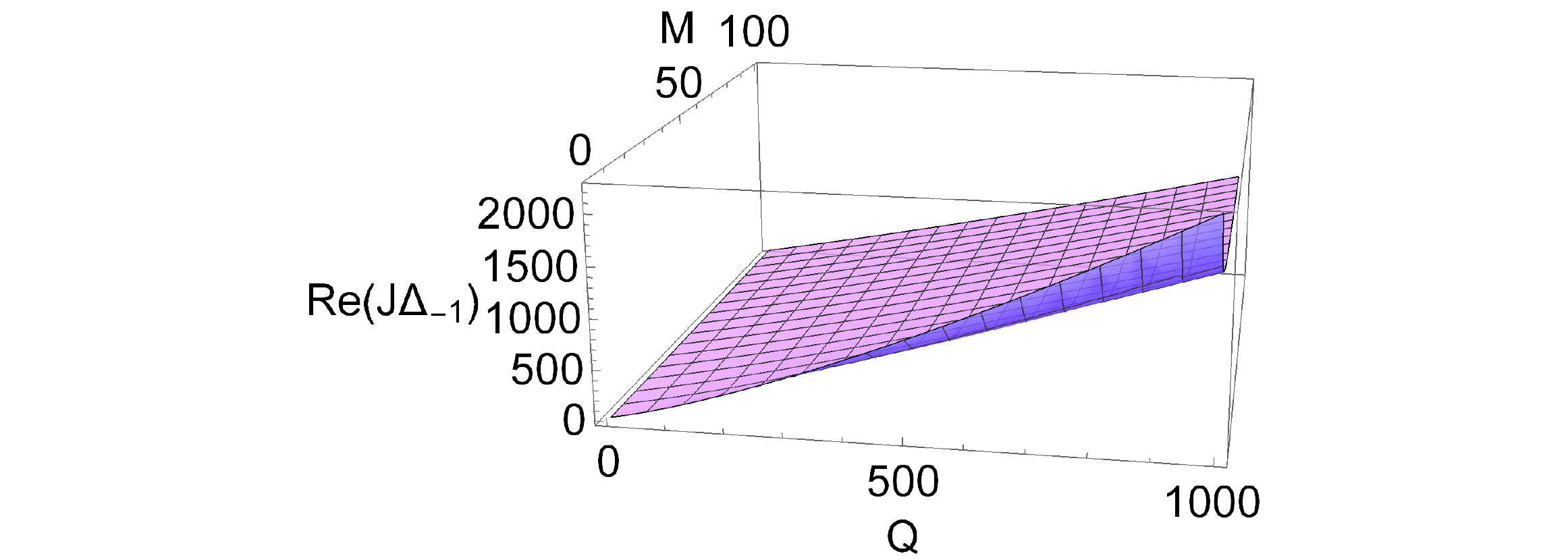}
	\caption{In this figure, we show the real part of the $U(N)\times U(M)$ model scaling dimension i.e.~$\rm{Re}\left(J\Delta_{-1}\right)$ as a function of charge $Q$ and variable $M$. We have chosen $N=4$, $s=1$ and $\epsilon=0.1$.}
	\label{UN_UM_Leading}
\end{figure}

\begin{figure}
\centering
	\includegraphics[width=0.75\columnwidth]{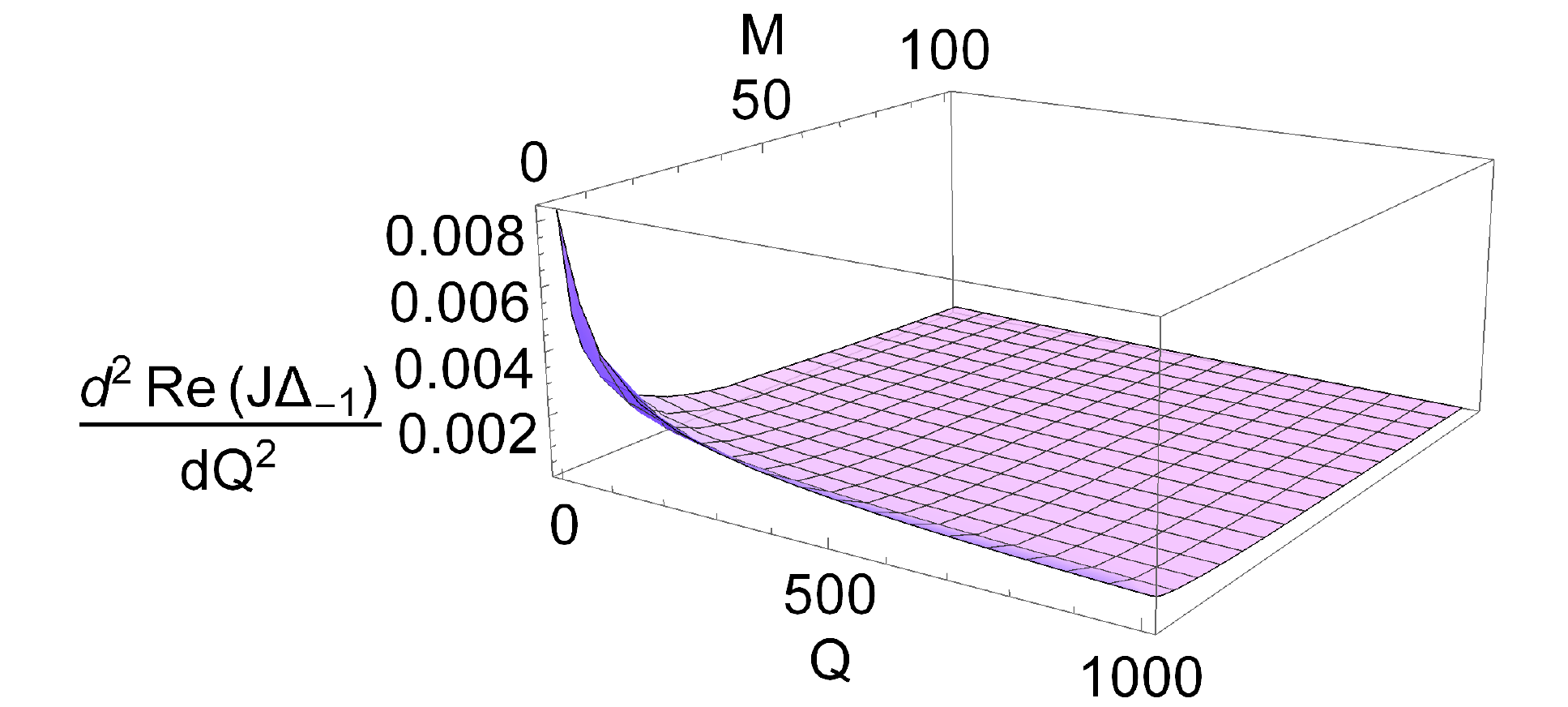}
	\caption{In this figure, we show the convexity property for the real part of $U(N)\times U(M)$ model scaling dimension i.e.~$\frac{d^2\rm{Re}\left(J\Delta_{-1}\right)}{dQ^2}$ as a function of charge $Q$ and variable $M$.  We have chosen $N=4$, $s=1$ and $\epsilon=0.1$.
	}
	\label{UN_UM_Leading_Curvature}
\end{figure}

The above quantity is positive when the couplings are positive. Plugging into $\gamma_{Q_1,Q_2}$ the expression for the FP couplings \eqref{complFP}, we have

\begin{align}
     \gamma_{Q_1,Q_2} &=\frac{Q_1 Q_2 \epsilon}{2 s \left((M N-8)
   (M+N)^2+108\right)} \Bigg(  \left(s
   \left(-M^2-2 M N-N^2+36\right) \right. \nonumber \\ & \left.+\sqrt{M^2-10
   M N+N^2+24} (s (M+N)-3)+M^2 N+M N^2-5 M-5
   N\right) \Bigg) + \cO\left(\epsilon^2\right)\,.
\end{align}
It is easy to check that, when the FPs are real, we have $ \gamma_{Q_1,Q_2} \ge 0$ and thus the convex charge conjecture, as formulated in Eq.\eqref{key_equation} holds. Furthermore, when the FPs are complex, the real part of $\gamma_{Q_1,Q_2}$ is still positive.
As the next step, we apply the semiclassical results \eqref{leading_UN_UM} and \eqref{NLO}
to study the convexity property. When getting into the regime where the fixed points of the couplings become complex, we also check the convexity of the real part of the scaling dimension.

In \autoref{UN_UM_Leading}, we present the real part of the leading order scaling dimension in the semiclassical expansion \eqref{leading_UN_UM}.  Using leading order semiclassical results (closed form) of Eq.~\eqref{leading_UN_UM}, we obtain $\rm{Re}\left(J\Delta_{-1}\right)$ as a function of charge $Q$ and variable $M$.
As reference values, we have chosen $N =4$, $s=1$ and $\epsilon=0.1$\footnote{We checked that the convexity property of the spectrum does not change if other values are chosen.}. The second derivative with respect to $Q$ is shown in \autoref{UN_UM_Leading_Curvature}. The convexity property is satisfied. For $N=4$, the fixed point values are complex below $M\sim40$ (see \autoref{real_complex_phase_diagram}). Interestingly, from \autoref{UN_UM_Leading_Curvature}, the convexity property  holds for the real part of the scaling dimension in the region of complex fixed points.

\begin{figure}[!t]
\centering
\begin{subfigure}{.5\textwidth}
  \centering
  \includegraphics[width=0.96\columnwidth]{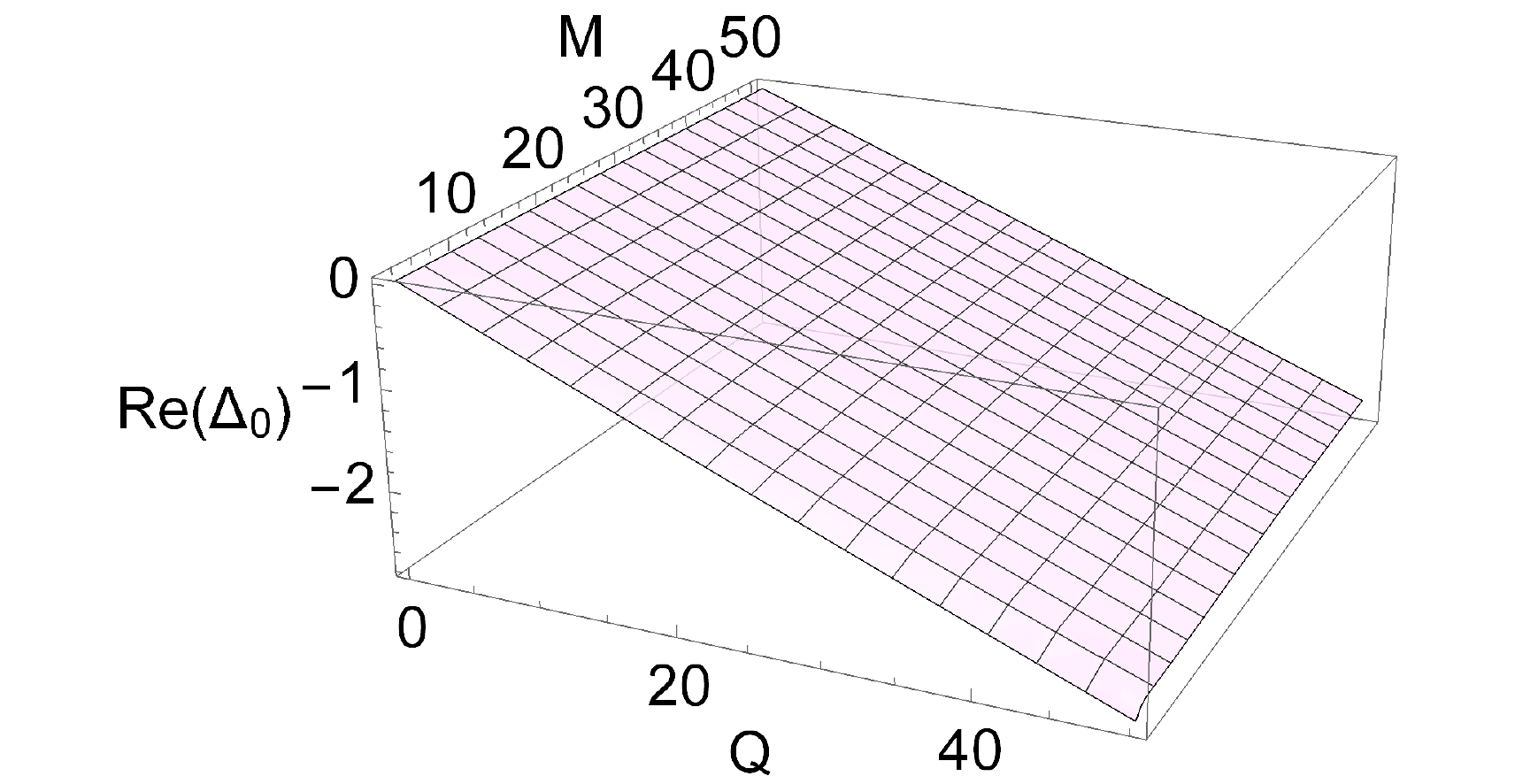}
\end{subfigure}%
\begin{subfigure}{.5\textwidth}
  \centering
  \includegraphics[width=1.07\columnwidth]{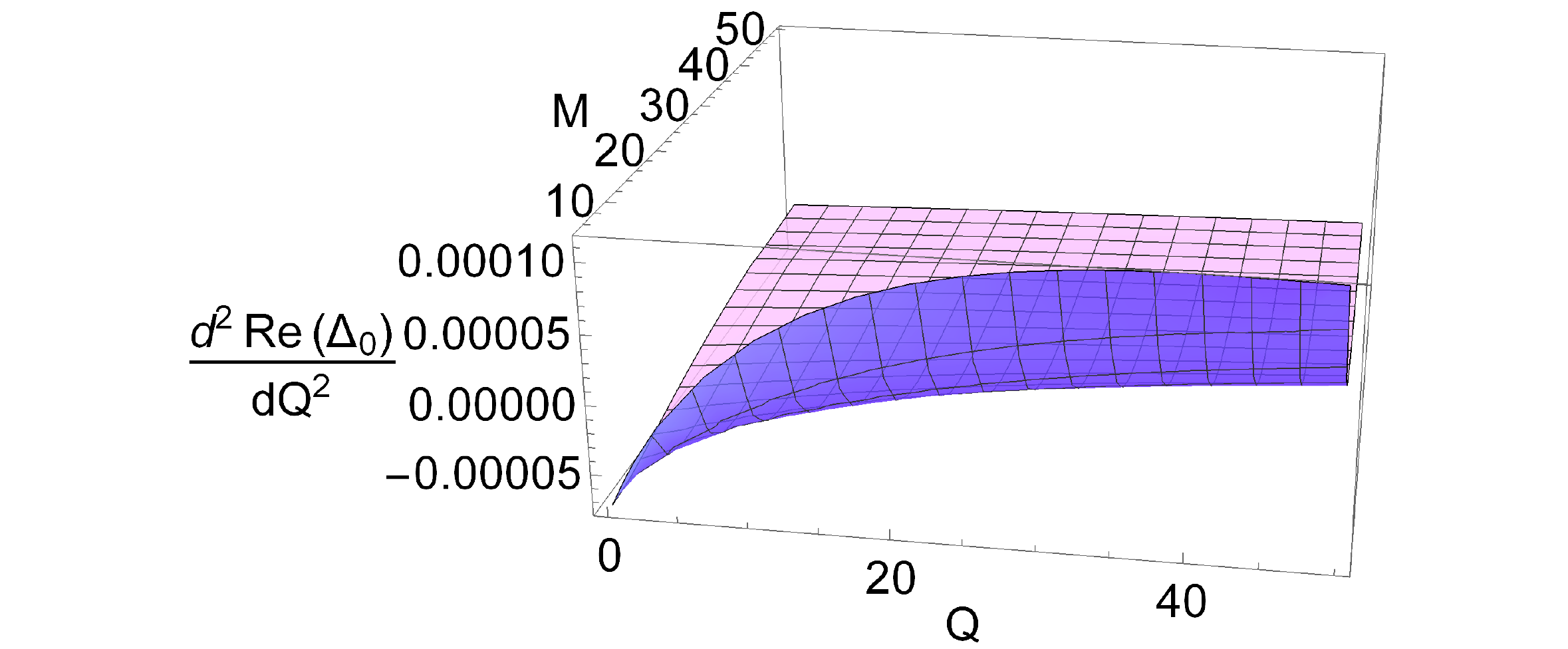}
\end{subfigure}
\caption{In the left figure, we show the real part of the $U(N)\times U(M)$ model scaling dimension at next to leading order i.e.~$\rm{Re}\left(\Delta_{0}\right)$ as a function of charge $Q$ and variable $M$ while in the right one, we show the convexity property for the real part of the of scaling dimension i.e.~$\frac{d^2\rm{Re}\left(\Delta_{0}\right)}{dQ^2}$ as a function of charge $Q$ and variable $M$. We have chosen $N=4$ and $s=1$ and $\epsilon=0.1$.}
\label{UN_UM_Next_Leading}
\end{figure}

In the left part of \autoref{UN_UM_Next_Leading}, using the semiclassical results of Eq.~\eqref{NLO}, we present the real part of the scaling dimension at next-to-leading order i.e.~$\rm{Re}\left(\Delta_{0}\right)$ as a function of charge $Q$ and variable $M$. In the right part of \autoref{UN_UM_Next_Leading}, we show its convexity property i.e.~$\frac{d^2\rm{Re}\left(\Delta_{0}\right)}{dQ^2}$ as a function of charge $Q$ and variable $M$. We have chosen $N=4$ charge configuration $s=1$ and $\epsilon=0.1$. Similar to the $O(N)$ case, the convexity property does not hold for the next-to-leading order in the small charge regime with very small 't Hooft coupling.


\begin{figure}
\centering
\begin{subfigure}{.5\textwidth}
  \centering
  \includegraphics[width=1.0\columnwidth]{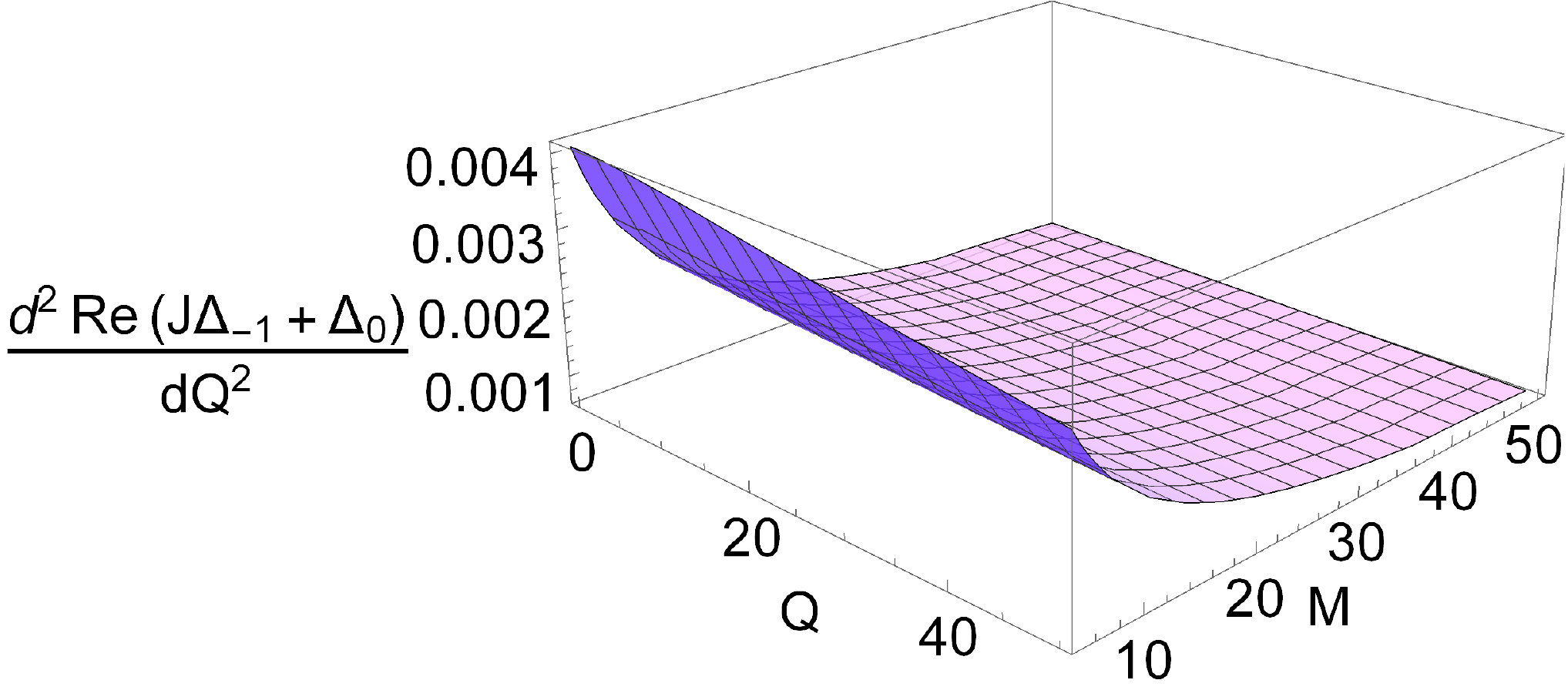}
\end{subfigure}%
\begin{subfigure}{.5\textwidth}
  \centering
  \includegraphics[width=0.9\columnwidth]{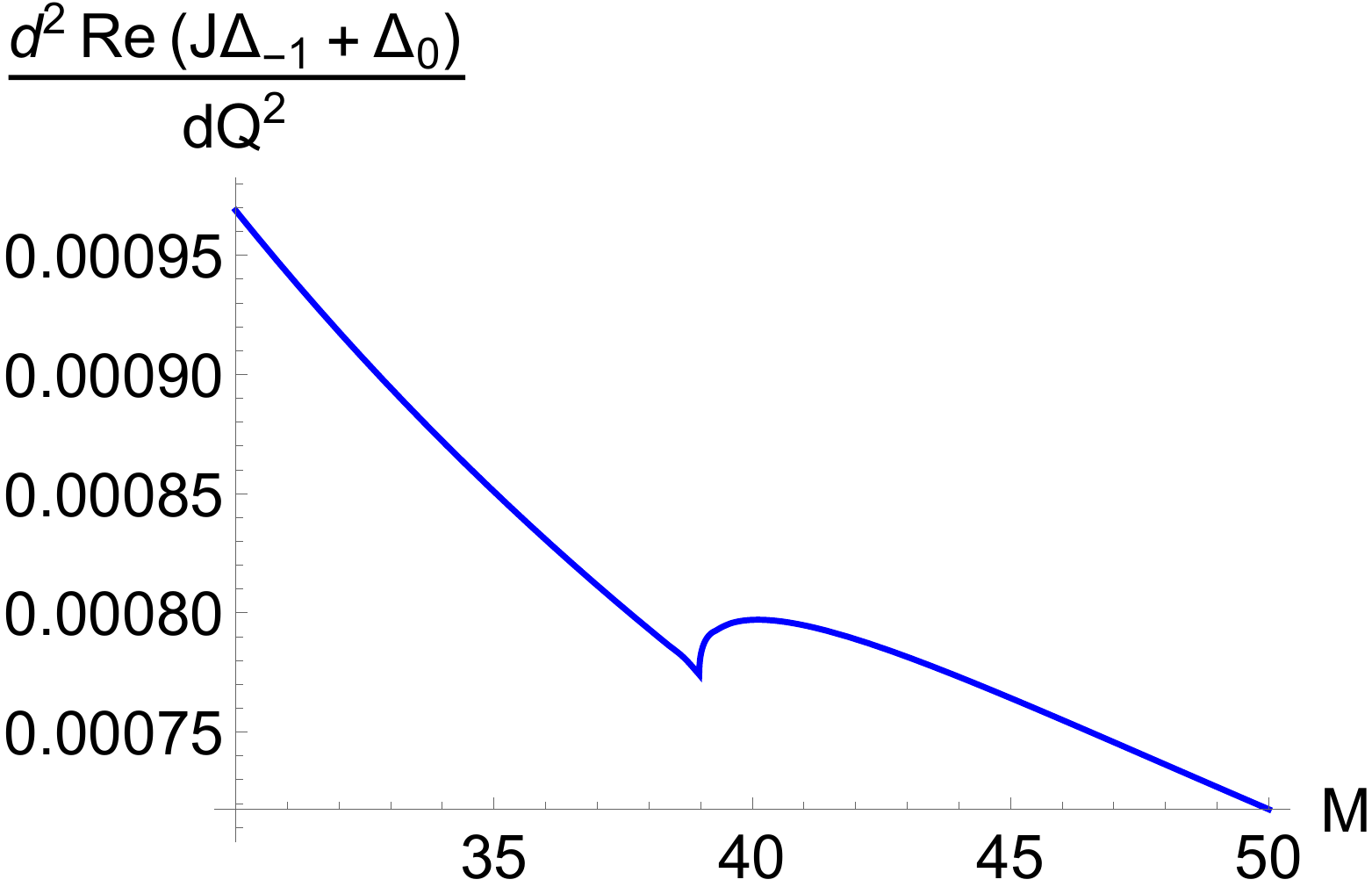}
\end{subfigure}
\caption{In the left figure, we show the convexity property for the real part of the $U(N)\times U(M)$ model leading plus next-to-leading order scaling dimension i.e.~$\frac{d^2\rm{Re}\left(J\Delta _{-1}+\Delta _0\right)}{dQ^2}$ as a function of charge $Q$ and variable $M$. We have chosen $N=4$, $s=1$ and $\epsilon=0.1$. In the right figure, we have chosen a slice of the left 3D figure with $Q=25$. The dip comes from the transition from the phase boundary from real to complex.}
\label{UN_UM_Leading+Next_Leading_Curvature}
\end{figure}

In \autoref{UN_UM_Leading+Next_Leading_Curvature}, we show the convexity property for the sum of real part of leading and next-to-leading order scaling dimension i.e.~$\frac{d^2\rm{Re}\left(J\Delta _{-1}+\Delta _0\right)}{dQ^2}$ as a function of charge $Q$ and variable $M$. We have chosen $N=4$, $s=1$ and $\epsilon=0.1$. This is an intriguing result, since it means the convexity property holds even in the non-unitary theory where the fixed points become complex if we focus on the real part of the scaling dimensions. On the right part of \autoref{UN_UM_Leading+Next_Leading_Curvature}, we have chosen a slice which shows $\frac{d^2\rm{Re}\left(J\Delta _{-1}+\Delta _0\right)}{dQ^2}$ is continuous with variable $M$ though there is a dip across the phase boundary from real to complex.

\begin{figure}
\centering
	\includegraphics[width=0.6\columnwidth]{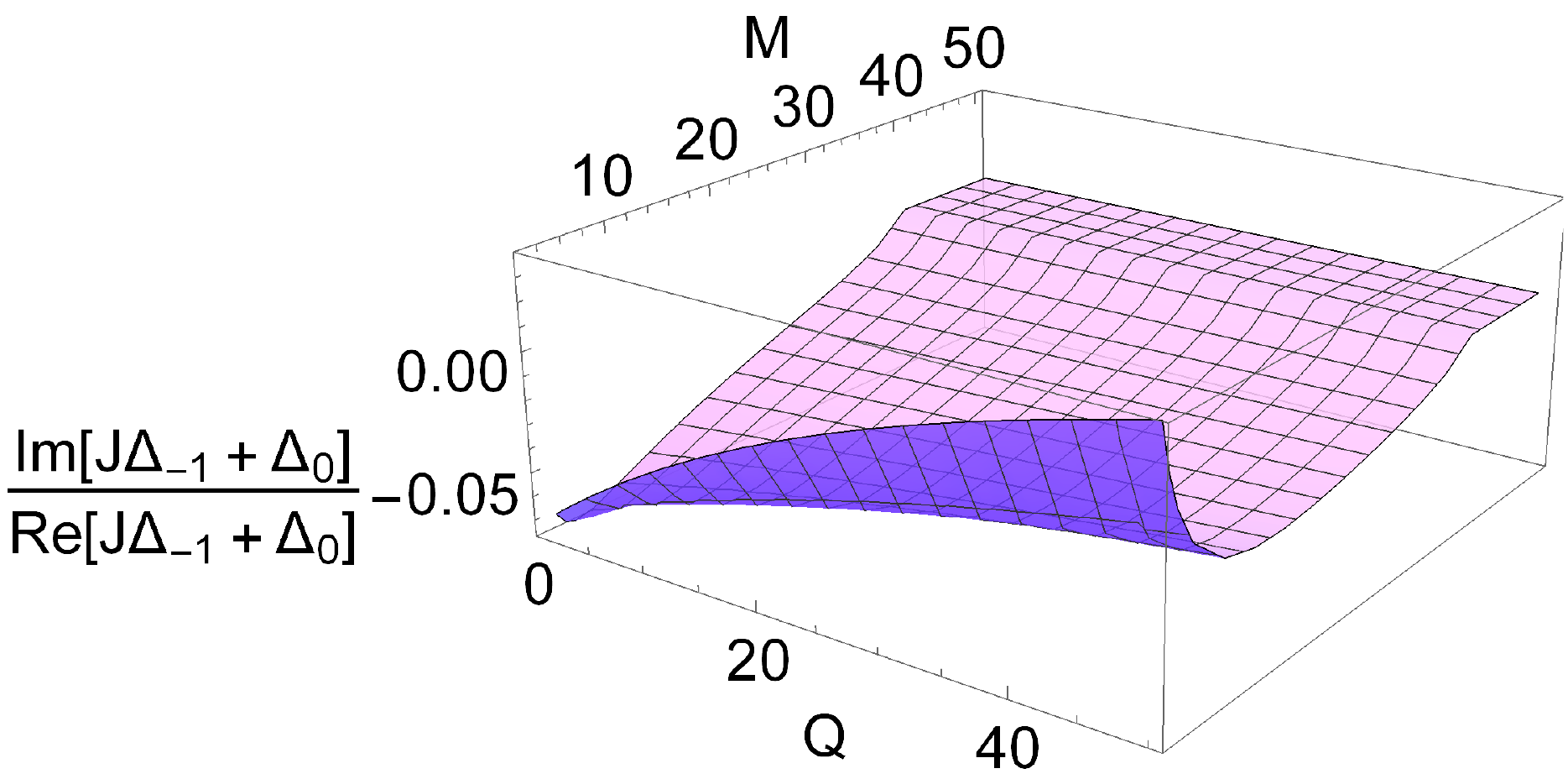}
	\caption{In this figure, we show the ratio of the imaginary and real part of the $U(N)\times U(M)$ model scaling dimension i.e.~$\frac{\rm{Im}\left(J\Delta _{-1}+\Delta _0\right)}{\rm{Re}\left(J\Delta _{-1}+\Delta _0\right)}$ as a function of charge $Q$ and variable $M$. We have chosen $N=4$, $s=1$ and $\epsilon=0.1$.}
	\label{UN_UM_Ratio_Imaginary_Real}
\end{figure}

In \autoref{UN_UM_Ratio_Imaginary_Real}, we present the ratio of imaginary and real part of the $U(N)\times U(M)$ model scaling dimension i.e.~$\frac{\rm{Im}\left(J\Delta _{-1}+\Delta _0\right)}{\rm{Re}\left(J\Delta _{-1}+\Delta _0\right)}$ as a function of charge $Q$ and variable $M$. We have chosen $N=4$, $s=1$ and $\epsilon=0.1$.

\bigskip
\textbf{Charge Configuration Two:}\\
In \cite{Antipin:2021akb}, we considered one more charge configuration given by
\begin{align} \label{seconda}
\mathcal{Q}_J=\rm{diag}\left\{-2J,J,J,0,\cdots,0\right\} \,.
\end{align}
In this case, the expansion for small 't Hooft couplings of the leading order of the semiclassical expansion \eqref{forma}, reads
\begin{align}
J \Delta_{-1} & = Q\left[ 1+\frac{ Q (3 u^*+8 v^*)}{64
   \pi ^2} - \frac{Q^2 \left(5 u^{*2}+24 u^* v^* + 32 v^{*2}\right)}{1024 \pi ^4} + \cO\left((u^* Q)^3, (v^* Q)^3\right) \right] \,.
\end{align}
where we set the couplings to their FP values and $Q = 8 J$ is the classical scaling dimension of the corresponding lowest-lying operator.
Using the above result, we have
\begin{equation}
    \gamma_{Q_1,Q_2}=\frac{Q_1 Q_2}{32 \pi^2}(3 u^* + 8 v^*) + \cO\left(\epsilon^2\right)\,.
\end{equation}
Analogously to the previous charge configuration, $\gamma_{Q_1,Q_2}$ is positive if the couplings are positive. As shown before, this condition is satisfied for all the values of $N$ and $M$ leading to real couplings, validating the convexity conjecture. Furthermore, when the couplings are complex, their real part is always positive, leading to $\text{Re}[\gamma_{Q_1,Q_2}] >0$.

\begin{figure}
\centering
	\includegraphics[width=0.6\columnwidth]{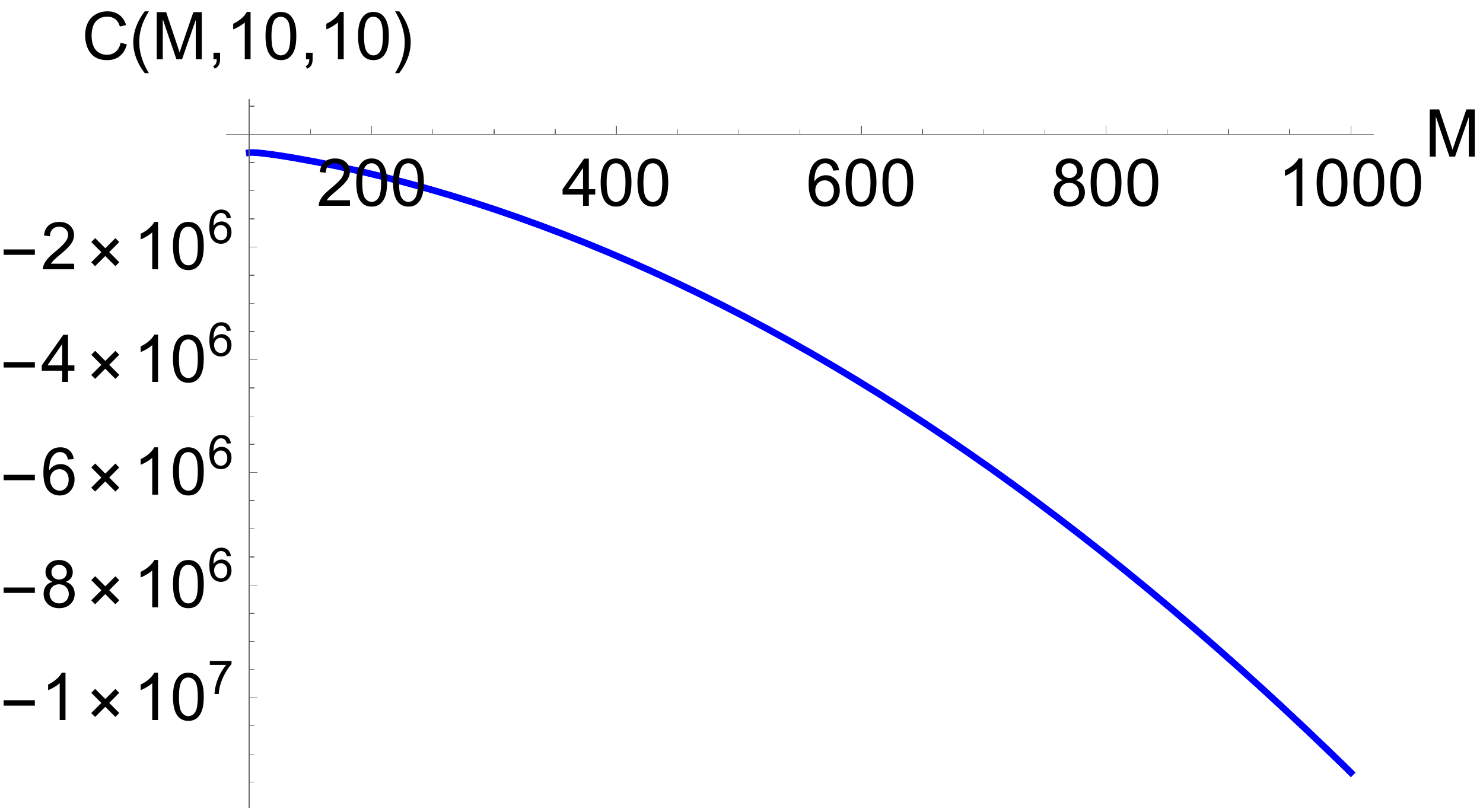}
	\caption{$C(M,N,J)$ for $N=J=10$ as a function of $M$. Since $C(M,N,J)<0$ the generalize version of the conjecture stated in \eqref{genconvex} is violated.}
	\label{UN_UM_newconjecture}
\end{figure}

We conclude this section with an example which illustrates that the generalized convex charge conjecture \eqref{genconvex} is not satisfied by the spectrum of the $U(N)\times U(M)$ CFT. In fact, by considering the family of charge configurations given in Eq.\eqref{complete}, we can re-state the conjecture as
\begin{equation}
    \gamma_{s_1,s_2} \equiv \Delta_{s_1 + s_2} -\Delta_{s_1} -\Delta_{s_2} \ge 0\,.
\end{equation}
On the other hand, by using the one-loop result in \eqref{one-loop}, we have
\begin{equation}
       \gamma_{s_1,s_2} = \frac{J s_1 s_2 (4 J v^*-u^*)}{\pi ^2} + \cO\left( \epsilon^2 \right) = \frac{4 J s_1 s_2  \mathcal{C}(M,N,J)
  }{(M N-8) (M+N)^2+108} \epsilon + \cO\left( \epsilon^2 \right)\,,
\end{equation}
where
\begin{align}
C(M,N,J) &= 2 J \left((M+N) \left(\sqrt{M^2-10 M
   N+N^2+24}-M-N\right)+36\right) \\ \nonumber & +3
   \sqrt{M^2-10 M N+N^2+24}+M (-N) (M+N)+5 M+5
   N\,.
\end{align}
The sign of $ \gamma_{s_1,s_2}$ is equal to the one of $C(M,N,J)$. It is easy to check that $C(M,N,J) < 0$, violating the generalized version of the convex charge conjecture. To illustrate this fact with an example, in Fig.\ref{UN_UM_newconjecture} we plot $C(M,N,J)$ for the reference values $N=J=10$, in a range of values of $M$ such that the FPs are real.

\section{Discussions and Conclusions}
Triggered by the recently introduced convex charge conjecture, related to the weak gravity one, we first considered different mathematical definitions and then we thoroughly tested them against a number of CFTs at nonzero charge. Our methodology of choice is the semiclassical framework that is eminently suited for analyzing fixed charge sectors of different theories in various dimensions. Concretely, we analyzed the convexity properties of the leading and next-to-leading order terms stemming from  semiclassical computations. We were therefore able to extend and go beyond the original tests of the conjecture. We analyzed the convexity of the fixed charge conformal dimensions to the leading  and next-to-leading order in the semiclassical framework.  Both contributions are known in a close form and for arbitrary charges and to all orders in the fixed point coupling.  Although the leading contribution is sufficient to test convexity in the semiclassical computations, we found instructive to determine the convexity property of the next-to-leading order as well. For the unitary theories investigated here the full semiclassical leading order is always convex while the next-to-leading order is concave for small values of $\lambda_\ast Q$.

 The situation is much more intriguing for models featuring a transition to complex conformal dimensions either as function of the charge or number of matter fields. The $O(N)$ model in $4+\epsilon$ dimensions is an example of the first kind. Here, in agreement with \cite{Aharony:2021mpc}, we have that for small charges the conformal dimension is real but concave rather than convex because the theory does not have a ground state. Intriguingly, as we increase the charge also an imaginary component emerges. The real part continuously connects, as function of the charge, with the one at small charge but changes character (meaning that the first derivative is discontinuous) and becomes convex.  The matter field example is represented by the $U(N)\times U(M)$ model in $4-\epsilon$ dimensions. This model displays an interesting phase structure because we can go from real to complex conformal dimensions by changing the number of matter fields. We observe that within the phase diagram in which  the conformal dimensions are real, convexity holds. Once the imaginary part develops the real part is still convex and continuously connected to the zero imaginary part one.

\section*{Acknowledgements}
We thank Ofer Aharony and Eran Palti for relevant discussions. The work of O.A. and J.B. is partially supported by the Croatian Science Foundation project number 4418. F.S and Z.W acknowledge the partial support by Danish National Research Foundation grant DNRF:90.
C.Z. is supported by MIUR under grant number 2017L5W2PT and INFN grant STRONG.

\bibliography{wgc_v2}

\end{document}